\documentclass[11pt]{article}
\usepackage[xdvi]{graphicx}
\usepackage{epsfig}
\usepackage{amsfonts, amssymb, amsmath}
\usepackage{subfigure}
\usepackage{latexsym,a4wide}

\newcommand{\be}{\begin{equation}}
\newcommand{\ee}{\end{equation}}
\newcommand{\ba}{\begin{eqnarray}}
\newcommand{\ea}{\end{eqnarray}}
\newcommand{\ban}{\begin{eqnarray*}}
\newcommand{\ean}{\end{eqnarray*}}
\newcommand{\bml}{\begin{mathletters}}
\newcommand{\eml}{\end{mathletters}}

\font\cat=cmr7
\def\P{\cat{P}}

\def\half{\textstyle{1\over2}}

\def\quarter{\textstyle{1\over4}}

\def\ie{{\it ie }}
\def\M{{\cal{M}}}
\def\L{{\cal{L}}}
\def\A{{\cal A}}

\def\L{{\cal L}}
\def\E{\cat{E}}
\def\M{{\cal M}}
\def\Q{{\cal Q}}

\def\P{\cat{P}}

\def\T0{{\bar{ T_0}}}
\def\n{{\mathrm{n}}}
\def\N{{\cal{N}}}
\def\E{{\cal{E}}}
\def\k{{\cal{K}}}
\def\R{{\cal{R}}}

\newcommand{\dz}{\partial_z}
\newcommand{\mpl}{M_\mathrm{Pl}}
\newcommand{\sgn}{\, \mathrm{sgn}\,}

\def\pp{/ \hspace {-2pt} /}

\begin{document}

\title{Higher order gravity theories and their black hole solutions}

\author{Christos Charmousis, \\ {\it LPT, Universit\'e de Paris-Sud,}\\
{\it B\^at. 210, 91405 Orsay CEDEX, France}}

\maketitle

\begin{abstract}
In these lectures notes, we will discuss a particular higher order gravity theory, Lovelock theory, that generalises in higher dimensions than 4, general relativity. After briefly motivating modifications of gravity, we will introduce the theory in question and we will argue that it is a unique, mathematically sensible, and physically interesting extension of general relativity. We will see, by using the formalism of differential forms, the relation of Lovelock gravity to differential geometry and topology of even dimensional manifolds.
We will then discuss a generic staticity theorem, quite similar to Birkhoff's theorem in general relativity, which will give us the charged static black hole solutions. We will examine their asymptotic behavior, analyse  their horizon structure and briefly their thermodynamics. For the thermodynamics we will give a geometric justification of why the usual entropy-area relation is broken.  We will then examine the distributional matching conditions for Lovelock theory. We will see how induced 4 dimensional Einstein-Hilbert terms result on the brane geometry from the higher order Lovelock terms. With the junction conditions at hand, we will go back to the black hole solutions and give applications for braneworlds: perturbations of codimension 1 braneworlds and the exact solution for  braneworld cosmology as well as the determination of maximally symmetric codimension 2 braneworlds. In both cases, the staticity theorem evoked beforehand will give us the general solution for braneworld cosmology in codimension 1 and maximal symmetry warped branes of codimension 2.  We will then end with a discussion of the simplest Kaluza-Klein reduction of Lovelock theory to a 4 dimensional vector-scalar-tensor theory which has the unique property of retaining  second order field equations. We will comment briefly, the non-linear generalisation of Maxwell's  theory and scalar-tensor theory.  We will conclude by listing some open problems and common difficulties.

\textbf{Report.no\/}: LPT-08-42
\end{abstract}

\newpage

\tableofcontents


\section{An introduction to Lovelock gravity}

\subsection{Modifying general relativity}

A convenient starting point for treating modifications of gravity are the fundamental building 
blocks of general relativity (GR) itself.
According to Einstein's theory, gravitational interactions are 
described on a spacetime manifold by a symmetric metric tensor $g$ 
endowed with a metric  and torsion-free connection (by definition a Levi-Civita connection) 
that obeys Einstein's field equations. In component language these equations read,
\be
\label{chaeinstein}
G_{ab} + \Lambda_{bare} g_{ab}= 8 \pi G T_{ab}
\ee
where the Einstein tensor $G_{ab}=R_{ab}-\frac{1}{2} g_{ab} R$ is given with respect to the Ricci 
curvature tensor $R_{ab}$ and we have included $\Lambda$, the cosmological constant,
 and $T_{\mu\nu}$  the energy-momentum tensor. The field equations are acquired from the Einstein-Hilbert action,
\be
\label{chaeinhil}
S=\frac{1}{16 \pi G}\int_\M d^4x \sqrt{-g}\; \L(\M,g,\nabla)
\ee 
where the Langrangian is the functional
\be
\L(\M,g,\nabla)=-2\Lambda_{bare} +R,
\ee
by variation with respect to the metric $g$ and adequate boundary conditions (see for example the appendix in \cite{chawald}). The bare cosmological constant $\Lambda$ is a free parameter of the theory.

We expect Einstein's theory to break down at very high energies close to the Planck scale,
 $m_{Pl}^2=\frac{1}{16\pi G}$, 
where higher order curvature terms can no longer be neglected. Theories such as string theory or quantum loop gravity or again 
models, of extra dimensions, consider or model the effect of such modifications. 
GR on the other hand is very well tested at the solar system and by binary pulsar 
data in the regime of weak and strong gravity respectively \cite{chawill}. 
However, recent cosmological experiments, or astrophysical data, such as galactic rotation curves, 
or even the Pioneer anomaly, appearing just beyond solar system scales, could question the validity of GR even at classical scales at large enough distances. 
In particular, recent cosmological evidence, coming essentially from type Ia supernovae 
explosions \cite{chasup}, point towards an {\it actually} accelerating universe. 
Looking at (\ref{chaeinstein}) there are 3 theoretical directions one could pursue in order to interpret this result. 
Firstly, we can postulate the existence of an extremenly small positive cosmological constant of value, 
$\Lambda_{now} \sim (10^{-3}eV)^4$, fixed by the actual Hubble horizon size, driving the acceleration in (\ref{chaeinstein}). 
To get an idea of how tiny this constant is note that this minute energy scale
is most closely associated to the mass scale of neutrinos, $10^{-3} eV$. 
Hence, although such a possibility{\footnote{It is not an explanation until we find a precise mechanism of why it is there
at all and why now.}} is the most economic of all, since we 
can fit actual multiple data with the use of a single parameter, it actually demands an enormous amount of fine-tuning. Indeed, 
from particle physics, the vacuum energy contributions to the total value of the 
cosmological constant are of the order of the ultraviolet 
cut-off we impose to the QFT in question.
It can therefore range as far up  and close to the Planck scale
(for discussions on the cosmological constant problem and ways to explain it
see \cite{chaconstant}). The "big" cosmological constant problem is precicely how
all these vacuum energies associated to the GUT, SUSY, the standard model etc are fined-tuned each time to 0 by an exactly 
opposite in value bare cosmological
 constant $\Lambda_{bare}$ appearing in (\ref{chaeinhil}).
The unexplained small value of the cosmological constant $\Lambda_{now}$ 
is then an additional two problems to add to the usual "big" cosmological constant problem, 
namely, why the cosmological constant is not cancelled exactly to zero and why do we observe it now. 

A second alternative explanation one can consider, is that the accelereted expansion is due to a cosmological 
fluid of as yet unknown matter, dubbed dark energy, such as a quintessence (scalar) field with some potential 
 appearing in the right hand side of (\ref{chaeinhil}). 
One of the basic strengths of this approach is its simplicity and in some cases an interesting approach to the cosmological coincidence problem. Among its basic 
weaknesses, apart from the usual generic fine-tuning and 
stability problems to radiative corrections, 
is that if we sum-up the as yet undiscovered  matter sectors of the Universe ie, dark matter and dark energy, 
we conclude that only a mere 4\% percent  of the actual matter that constitutes our universe in its actual state has been discovered in ground based accelerators!
Although there exist theoretically motivated dark matter candidates, such as neutralinos or axions, 
stemming from well motivated particle theories, our understanding of dark energy, is rather poor. 
A third, far more ambitious alternative, that is less well studied, far more constrained, 
and admittedly less succesful up to now, 
is to modify the dynamics of geometry on the left hand side of (\ref{chaeinhil}) not only in the UV but also in the IR sector. 
This then would mean that Einstein's theory is also modified at large distances at the scale of the inverse 
Hubble scale of today as measured in a LFRW universe ($H_0/c=7.566 \times 10^{-27} m^{-1}$). 
This distance scale is enormous, to get an idea if we consider as our unit the distance of the earth to the sun (1 AU) 
we get {\footnote{ Astronomical units are interesting since most tests of General relativity are at distance scales of the 
solar system. Hence extrapolation to $10^{15}$ scales bigger of such experiments 
can be sometimes unjustified or at least questionable.}} a present horizon distance of $10^{15} AU$!  
 
Next question is how do we modify gravity consistently? 
 One can consider three basic types of modification which at the end of the day are not completely unrelated. 
Indeed, we can include additional fields or degrees of freedom, 
for example, scalar or vector, see for example  \cite{chabd}),  \cite{chajacob}),  we can 
enlarge the parameter space where the theory evolves, for example the number of dimensions, 
the geometric connection in question (we include torsion etc) or again we can generalise the field equations. 

In all cases, it is very important to fix  basic consistency requirements for the modified gravity theory. To fix the discussion we can ask for three basic requirements: 
first we would like that the theory under consideration be consistent theoretically, for example we ask for 
 sensible vacua of maximal symmetry, such as Minkowski, de Sitter or anti de Sitter spacetime, and  
valid stable perturbation theory around these vacua. We secondly need to satisfy all actual experimental constraints as
for ordinary GR plus we need correct IR cosmological behaviour without the need for dark energy nor a
  cosmological constant. We thirdly want our theory to have the least number of degrees of freedom possible 
  and to be naturally connect to GR theory{\footnote{The first and third requirements are not 
absolute but one needs to be aware at least when a theory does not validate one of these.}}. For example, Brans-Dicke theory \cite{chabd} clearly passes the first
   and third tests whereas  solar system constraints are rather restrictive \cite{chawill}. 

\subsection{Lovelock's theory}

In these notes we will  restrict our attention to a metric modification of gravity that generalises GR in higher dimensions. 
Remaining tangential to GR (principles) we consider a theory $\L=\L(M,g,\nabla)$, whose field variable is a single symmetric metric tensor $g$ 
endowed with a Levi-Civita connection $\nabla$. We ask for a divergence free geometric operator on the right hand 
 side of (\ref{chaeinhil}), since we know that matter obeys the conservation equation $\nabla^{\mu} T_{\mu\nu}=0$.
Furthermore, in order to bypass perturbative stability constraints for the graviton, we ask for second 
order field equations. These two properties are quite natural for our theory  if we want to extend GR
 at the classical level but, we emphasize, not necessary 
   at ultra-violet scales. Although higher derivatives generically introduce ghost degrees of 
freedom\cite{chawood} {\footnote{An exception to 
  this rule are $f(R)$ theories  \cite{chafr1} since they involve only functionals of the Ricci scalar. 
These theories have been known since a long time to be conformally equivalent to scalar-tensor theories see for example \cite{chafr}}} 
around the vacuum, \cite{chazwei} one may argue that these may dissappear 
having correctly summed the infinite number of higher order corrections. This is precicely the case in string theory
 which although is a ghost free theory of 2-dimensional surfaces embedded in 10 dimensions, 
at the effective action level, acquires (unphysical) ghost
 degrees of freedom because of the effective cut-off we impose. 
They are in general cured by arranging for the appearence of the relevant Lovelock 
term \cite{chagross} to the relevant order. 

In $D=4$ the only two derivative metric modification to Einstein's theory is the addition of a cosmological constant term! In other words, any higher order curvature invariant either  gives a pure divergence term, not contributing to the field equations, or, adds higher order derivatives to the field equations.
In higher dimensions this no-go extension theorem to GR is no longer true. It was the object of  Lovelock's theorem \cite{chalov}, (see \cite{chalan} for the $D=5$ case) to prove back in the seventies that there exist theories containing, precise higher order curvature invariants, that actually modify Einstein's field equations (\ref{chaeinhil}) while satisfying $\nabla^{\mu} T_{\mu\nu}=0$, in the face of modified Bianchi identities, and while keeping the order of the field equations down to second order in derivatives. The theory in question will be the subject of this brief study and gives in $D=4$ precicely GR with a cosmological constant and in 5 and 6 dimensions reduces to Einstein-Gauss-Bonnet theory (EGB). Lovelock theory in a nutshell is the generalisation of general relativity in higher dimensions while keeping the full generality of GR in $D=4$.

Following  Lovelock's proof  of the uniqueness theorem \cite{chalov} (see also the neat derivation of \cite{chazum} using differential forms) significant interest developped in these higher dimensional relativity theories in the eighties with motivations originating from string theory \cite{chaant} and others originating  from Kaluza-Klein cosmology \cite{chamad}. Initial interest in string theory was trigerred by  Zwiebach \cite{chazwei} who noted that second order corrections to the Einstein-Hilbert action, other than the Gauss-Bonnet invariant, introduced a graviton ghost when considering perturbations around  flat spacetime. Effective action calculations of certain string theories  \cite{chagross} found that the leading (tree-level in $g_S$) $\alpha'$ string tension corrections could give rise, modulo field redifinitions to this order, to the Gauss-Bonnet invariant. Several nice papers appeared uncovering properties and analysing exact solutions of EGB  \cite{chaboul}, \cite{chawheeler}, \cite{chawiltshire0} while slightly later  tackling full Lovelock theory exact solutions  \cite{chams}. More recently there have been a few exact solutions discussed \cite{chaman}, \cite{chadeh} and some solution generating techniques developped (see for example \cite{chakastor}). Discussions on issues of energy, stability and the hamiltonian formalism have been carried out in \cite{chatek}, \cite{chakofi3} \cite{chapad2}.

Interest in Lovelock theory and in particular its 5 and 6 dimensional version, EGB theory, has attracted quite a lot of attention recently in the context of braneworlds (see Ruth Gregory's lecture notes \cite{charuth0}). Indeed from the braneworld point of view it would seem important to consider the general bulk theory rather than just GR in 5 or 6 dimensions and investigate if the 4 dimensional braneworld picture remained GR like. In a nutshell (we will uncover the details later on) the Gauss-Bonnet  term in the bulk  action is  similar in nature as is the induced gravity term \cite{hol}, \cite{chadgp} to be added to the brane action. Loosely speaking, it thus enhances GR type effects on the brane adding also quite naturally  a UV modification  to the usual one identified by the 5th dimension. Perturbation theory in the bulk is exactly the same around a maximally symmetric spacetime and the main difference are the boundary conditions on the brane which become mixed \cite{chafax2}, similarily for those for induced gravity. In order to evaluate the correct boundary conditions, that give the braneworld gravitational spectrum, and hence determine the 4 dimensional gravity and stability of the setup, the important difficulty one has to face, is finding the extension of the Israel junction conditions \cite{chaisr} in the context of EGB theory. In fact the junction conditions can be calculated directly, for each solution in question, by a careful calculation of the distributional terms{\footnote{As it was pointed out first in \cite{chadol} distribution theory does not allow for ordinary multiplication and this can lead to erronious junction conditions.}}  \cite{chamav}, \cite{chamei}, \cite{chajak} and \cite{chafax1} in the context of braneworld cosmology (see also \cite{chader}). The full covariant solution to the problem was first found by Davis, Gravannis and Willisson \cite{chadav} where it was realised that careful variation of the bulk metric with respect to the boundary term to this theory, discovered by Myers back in the eighties \cite{chamye2}, would give rise to the correct junction conditions. This is exactly similar to what happens when one considers careful variation of the Gibbons-Hawking boundary term thus obtaining Israel's junction conditions. Using the junction  conditions it was found that negative tension branes induced tachyonnic instabilities to braneworlds \cite{chafax2} as well as important changes in the tensor perturbation amplitudes for braneworld inflation \cite{chaduf}. Braneworld cosmology was further studied  with  particular focus in inflation \cite{chalid} and IR modifications \cite{chakof2}, \cite{chapad2}. For codimension 2 braneworlds the relevant matching conditions were shown to give \cite{chabos} precicely induced gravity terms on the brane plus extrinsic curvature corrections. However, up to now exact solutions or braneworld cosmology have not been found. We will discuss briefly here how one can obtain the maximally symmetric braneworld solutions in the context of EGB \cite{chaantonis}. The full matching conditions of Lovelock theory, irrespective of codimension  were given in covariant formalism in \cite{chazeg} and recentl maximally symmetric braneworld examples to codimension 4 were found by Zegers \cite{chacod4}

In this review we will therefore study the basic properties and important characteristics of Lovelock theory. In the next section we will begin by introducing the theory in differential form language as it is the most adequate way to recognise its nice features and why it has unique properties. After 
this geometric parenthesis, we will study important exact solutions 
of this theory concentrating on the case of static black holes and solitons. We will see that a generalised version of Birkhoff's theorem holds as for GR, and we will then analyse the static black hole solutions, their thermodynamics and the solitonic solutions. After this we will discuss matching conditions of Lovelock theories and we will see that in this aspect Lovelock's theory is in principle a far richer extension to Einstein's theory in higher dimensions. Having done this we will discuss braneworld applications in codimension 1 and codimension 2. We will close off by looking at the Kaluza-Klein reduction of these theories and the type of scalar-vector-tensor theories they predict \cite{chamul}.

\subsection{Basic definitions for Lovelock theory-differential form language}

Our aim in this section is to construct the higher order curvature densities which will be the building blocks of Lovelock theory and explain what makes them special. Indeed, in component language these will turn out to be precise but seemingly ad-hoc linear combinations or powers of the Riemann, Ricci tensor and Ricci scalar. We will thus use differential form language where we will see that they are indeed powers of the curvature 2 form with a precise and clear geometric interpretation (see also \cite{chamyers}, \cite{chamadnat}). 

Let $(\M,g,\nabla)$ denote a $D$-dimensional spacetime manifold $\M$ endowed with a smooth 
{\footnote{By smooth we mean here metrics of at least $C^2$ regularity. This will be relaxed to piecewise $C^1$ when we look at braneworlds, allowing for
 distributional matter sources.}} spacetime metric $g$. 
The connection $\nabla$ is taken to be a Levi-Civita connection. 
 To every point $P$ of spacetime $M$, we
associate a local orthonormal basis{\footnote{For a more complete account on the geometrical 
notions used here and precise examples see \cite{chaegu}}} of the tangent space $T_P M$,
$(e_A)$, with $A=1,...,D$ such that
\be
g(e_A,e_B)=\eta_{AB}
\ee
Equivelantly the metric can be expressed as 
$g=\eta_{AB} \theta^A\otimes \theta^B=g_{ab}dx^a dx^b $ where the 1-forms $\theta_A$ are precicely dual to the basis vectors $e_A$, since $\theta^A (e_B)=\delta^A_{\, B}$. 
The metric components in a coordinate frame $dx^a$ (we use small case latin letters for a coordinate frame and upper-case latin letters for an orthonormal frame) are thus $g_{ab}=\theta^A_a\theta^B_b \eta_{AB}$ where $\theta^A_a dx^a=\theta^A$. 
The dual 1-forms $\theta^A$ form a natural basis of the vector space of 
1-forms $\Omega^{(1)}(TM)$. 
In turn the antisymmetric
product of  1-forms $\theta^A$ can be used in order to construct 
a basis of the higher order forms acting  on
$TM$. For any $k$-form $w$ in $\Omega^{(k)}(TM)$, 
where $(0\leq k\leq D)$, can be written as,
\be
w=w_{A_1 \cdots A_k} \theta^{A_1} \wedge \cdots \wedge \theta^{A_k}
\ee
with $w_{A_1 \cdots A_k}$ some smooth function.  Following this simple recipe we can
define a (D-k)-form,
\be
\theta^{\star}_{A_1 \cdots A_k}=\frac{1}{(D-k)!} \epsilon_{A_1 \cdots A_k
A_{k+1} \cdots A_D} \theta^{A_{k+1}} \wedge \cdots \wedge \theta^{A_D}
\ee
where $\epsilon_{A_1 \cdots \cdots A_D}$ is totally antisymmetric in its
$D$ indices and $\epsilon_{12\cdots D}=1$. This quantity is called the Hodge
dual of the basis $\theta^{A_1} \wedge \cdots \wedge \theta^{A_k}$ of
$\Omega^{(k)}(TM)$. It defines a dual basis of forms in 
$\Omega^{(D-k)}(TM)$.
We can therefore write the Hodge dual of any k-form as,
\ba
\label{chahodge}
\star : \qquad \qquad \qquad \Omega^{(k)}(TM) \qquad \qquad &\rightarrow & \qquad \Omega^{(n-k)}(TM) \nonumber \\
\omega= \omega_{A_1 \cdots A_k} \theta^{A_1} \wedge \cdots \wedge
\theta^{A_k} \quad
&\rightarrow & \quad \star \omega = \omega^{A_1 \cdots A_k} \theta^{\star}_{A_1 \cdots A_k}
\ea
The wedge product of any form with its dual is a $D$-form,
which is by construction proportional to the volume element of spacetime, which we note as $\theta^\star$. Obviously for $k>D$ all forms are identically zero. A
useful identity is
\be
\label{chaident}
\theta^B\wedge
\theta^{\star}_{A_1...A_k}=\delta^B_{A_k}\theta^{\star}_{A_1...A_{k-1}}
-\delta^B_{A_{k-1}}\theta^{\star}_{A_1...A_{k-2}A_k}+ \cdots + (-1)^{k-1}\delta^B_{A_{1}}\theta^{\star}_{A_2...A_k}
\, .
\ee

Having constructed this tower of $k$-forms, for $0\leq k \leq D$ we now need define two quantities: the connection 1-form and the curvature 2-form.
The connection 1-form of $M$, $\omega^A{}_{B}$ which replaces the usual Christophel symbols in coordinate language, is defined by
\be
d\theta^A=-\omega^A{}_{B}\wedge \theta^B
\ee
since we have assumed a torsion-less connection. 
On the other
hand, the spacetime curvature 2-form is linked to the connection 
1-form via the (second Cartan structure) equation,
\be
\label{chacartan}
\R^A{}_{B}=d\omega^A{}_{ B} + \omega^A{}_{C} \wedge 
\omega^C{}_{ B} \, .
\ee
The ambient curvature 2-form is related to the Riemann  tensor components by
\be
\label{chacurv}
\R^A{}_{B}=\frac{1}{2}R^A{}_{BCD} \theta^C \wedge \theta^D \, ,
\ee
with respect to the spacetime Riemann tensor{\footnote{As a word of caution, the Riemann tensor components are given here with respect to the orthonormal basis and may differ, as is the case for stationary spacetimes for example, to coordinate basis components of the same tensor.}}. 

Langrangian densities appear under spacetime integrals; they are therefore by definition going to be $D$-forms. 
Hence in order to build such densities out of the curvature 2-form (\ref{chacurv})
and its powers, symbolically $\R^k$, we need to construct $D$-forms 
by correctly completing with the relevant Hodge duals (\ref{chahodge}). 
The higher the dimension of $\M$ the higher the possible powers of curvature we can consider.
In differential form language it is straightforward to build this way the k-th Lovelock lagrangian density
${\cal L}_{(k)}$ which is a $D$-form defined by,
\be
\label{chalk}
{\cal L}_{(k)}=\R^{A_1 B_1}\wedge \dots\wedge \R^{A_k B_k} \wedge
\theta^{\star}_{A_1B_1 \dots A_kB_k}= \bigwedge_{i=1}^k \R^{A_i B_i}
\wedge \theta^{\star}_{A_1B_1 \dots A_kB_k}
\ee
and we stress that $k$ stands for the power of curvature.
Clearly,  ${\cal L}_{(0)}$ is the volume element, giving rise to a cosmological constant 
whereas it is easy to check using (\ref{chaident}) that
\be
\label{chaein1}
{\cal L}_{(1)}= \R^{A_1 B_1} \wedge \theta^{\star}_{A_1B_1}=R \, \theta^\star
\, ,
\ee
is the Ricci scalar density and
\be
\label{chahat}
{\cal L}_{(2)}= \R^{A_1 B_1} \wedge  \R^{A_2 B_2} \wedge 
\theta^{\star}_{A_1B_1A_2B_2}=(R^{ABCD}R_{ABCD}-4R^{AB}R_{AB}+R^2)\, \theta^\star 
\ee
is the Gauss-Bonnet density which we will denote by $\hat{G}$. 
Note that for $k>D/2$, (\ref{chalk}) vanishes so that if $D=4$ say, then  
${\cal L}_{(0)}$, 
${\cal L}_{(1)}$ and ${\cal L}_{(2)}$ are the only terms
present in the action (although ${\cal L}_{(2)}$, 
as we will now see, turns out to be trivial).  
According to Lovelock's theorem \cite{chalov}, given a metric theory, ${\cal L}_{(k)}$ 
are the unique densities, made out of $(\M,\nabla,g)$ as defined in the beginning of this section, 
which allow for energy conservation and second order field equations. 

\subsection{Lovelock densities and their geometric interpretation}

So what is special about these densities? The answer lies in differential geometry.
Indeed, if $D$ is even,
for $k=D/2$, the Hodge dual is trivial and the Lovelock density $\L_{D/2}$ reduces to,
\be
\label{chaeuld}
{\mathcal L}_{(D/2)}= \bigwedge_{i=1}^{(D/2)} \R^{A_i B_i} \,
\epsilon_{A_1B_1 \dots A_{D/2}B_{D/2}} \, ,
\ee
This $D$-form can be recognised as the generalised  Euler density for an even dimensional compact manifold $\M$ \cite{chachern}. It is a geometric quantity whose integral over $\M$ is a topological invariant (see for example \cite{chaspivak})
\be
\chi \left [ \M \right ]= \frac{1}{(4\pi)^{D/2} (D/2)!} \int_{\M} {\cal
  L}_{(D/2)} \, .
\ee
This relation has  important consequences since it yields a relation between a geometric quantity involving curvature with the topology of the manifold $\M$. This relation is familiar for the case of $D=2$ where the $k=1$ density, ie the Ricci scalar or Gauss curvature, gives the usual Euler characteristic for 2 dimensional compact surfaces with no boundary. This familiar formula then relates surface geometry with a topological number defining topologically equivalent classes  of surfaces, 
\be
\chi \left [ \M \right ]= \frac{1}{(4\pi)} \int_{\M} R=2-2h
\ee
where $h$  denotes the number of handles of $\M$ (see figure 1). The above relation gives in essence the Gauss-Bonnet theorem. 
\begin{figure}
	\centering
	\includegraphics[bb=0 0 401 103]{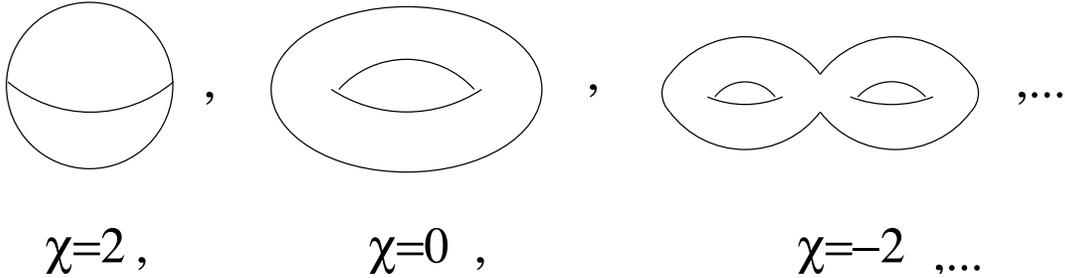}
	\caption{Any 2 dimensional compact surface can be continuisly deformed to one of its Euler classes
 parametrised by $\chi$.}
	\label{fig:2}
\end{figure}
This classification is also familiar from string theory where $\chi$ is related to the string 
coupling $g_s=e^{\chi \phi}$ giving the string surface diagrams (rather than Feynmann diagrams as for point particles) for orientable strings. 
Variation of this quantity with respect to the local frame $\theta^A$ and use of the Bianchi identities on the curvature 2-form, gives us  that ${\mathcal L}_{(D/2)}$ is locally an exact form. This in turn tells us that variation of (\ref{chaeuld}) gives no contribution to the field equations (supposing $\M$ compact).
In turn for $D=4$, ${\mathcal L}_{(2)}$ stands for the Gauss-Bonnet density whose
integral is now in turn the 4 dimensional Euler characteristic.

Therefore we see that the special feature of  Lovelock theory is that its Langrangian densities are {\it{dimensional continuations}} of the Chern-Euler forms{\footnote{By dimensional continuation we mean that we take $k$ powers of the curvature 2-form and conveniently "multiply" by the relevant Hodge dual as in (\ref{chalk}). }} which at lower even dimension are topological invariants. For completeness we note that for an even dimensional manifold with a boundary, the  Chern-Gauss-Bonnet  relation, is corrected by a Chern boundary form \cite{chachern}. 
\be
\label{eul}
\chi \left [ \M \right ]= \frac{1}{(4\pi)^{D/2} (D/2)!} \left [\int_{\M}
{\cal L}_{(D/2)} + \int_{\partial \M} \Phi_{(D/2-1))} \right ] \, ,
\ee
where the $(D-1)$-form $\Phi_{(D/2-1)}$ reads 
\cite{chaspivak,chachern,chaegu},
\be
\label{phi1}
\Phi_{(D/2-1)}= \sum_{m=0}^{D/2-1} \frac{D \cdot (D-2) \cdots 2(m+1)}{1\cdot
3 \cdots (D-2m-1)} \epsilon_{\mu_1 \cdots \mu_{D-1}} \left (\bigwedge_{l=0}^m
\R^{\mu_{2l-1} \mu_{2l}} \right )
\wedge \left ( \bigwedge_{l=0}^{D-1} \k^{\mu_l}{}_{N} \right ) \, .  
\ee
where $\k^{\mu_l}{}_{N}$ is a 1-form associated to the extrinsic curvature of the boundary $\partial \M$ \cite{chamyers}, \cite{chazeg}.
It turns out that the relevant boundary terms for Lovelock theory are again the dimensional continuations of the Chern boundary forms which are multiples of the induced Riemann tensor on the brane and extrinsic curvature of the boundary. This is a rather useful statement when treating braneworlds since variation with respect to the local frame gives the codimension 1 junction conditions \cite{chazeg} as we will see later on. The dimensionally continued boundary term associated to the Euler characteristic (\ref{chaein1}), $k=1$, is the Gibbons-Hawking boundary term \cite{chagib}
whereas the boundary term associated to the Gauss-Bonnet invariant is the Myers boundary term \cite{chamye2}. For details see \cite{chazeg}.

The integral of the sum, for $k<D/2$, of all Lovelock densities ${\cal
  L}_{(k)}$ is the most general classical action for $\M$, yielding up to
second order field equations for the metric tensor. This is the Lovelock
action, 
\be
\label{lla}
S_D= \int_{M} \sum_{k=0}^{[(D-1)/2]} \alpha_k {\cal L}_{(k)} \, ,
\ee
where the brackets stand for the integer part. 
The first three terms of this sum
\be
S_D = \int_{M} \left ( \alpha_0 \theta^{\star} + \alpha_1 \R^{AB}\wedge
\theta^{\star}_{AB} + \alpha_2 \R^{AB} \wedge \R^{CD}
\wedge \theta^{\star}_{ABCD} + \cdots \right )
\ee
are respectively the cosmological constant, Einstein-Hilbert and
Gauss-Bonnet terms, yielding the generalisation of the Einstein-Hilbert  action in
$D=5$ and $6$ dimensions. 

A variation of the action (\ref{lla}) including matter, 
with respect to the frame, gives the Lovelock equations
\be
\label{ll}
\sum_{k=0}^{[(D-1)/2]} \alpha_k {\cal E}_{(k) A} = -2T_{AB} \theta^{{\star} B}
\, ,
\ee
where ${\cal E}_{(k) A}$ is the k-th Lovelock (D-1)-form,
\be
\label{klov}
{\cal E}_{(k) A}= \bigwedge_{i=1}^k \R^{A_i B_i} \wedge 
\theta^{\star}_{AA_1B_1 \dots A_kB_k}
\ee
and we have chosen the normalisation according to 
$$
-\half \R^{A_1 B_1} \wedge \theta^{\star}_{CA_1B_1}=G^A_C \theta^{\star}_{A} \, ,
$$
so that the equations of motion read, in component formalism, $G_{AB} +\cdots
= T_{AB}$. For $k=2$ we get,
\be
{\cal E}_{(2) A}= \R^{A_1 B_1} \wedge   \R^{A_2 B_2}
\theta^{\star}_{AA_1B_1 A_2B_2}= H^A_C \theta^\star_C
\ee
where applying (\ref{chaident}) and (\ref{chacurv}) we obtain in component language,
\be
\label{chalov23}
H_{ab}={g_{ab}\over 2}
(R_{efcd}R^{efcd}
-4R_{cd}R^{cd}+R^2)\nonumber\\-2R
R_{ab}+4R_{ac}R^{c}_{\mbox{ }b} 
+4R_{cd}R_{\mbox{ }a\mbox{ }b}^{c\mbox{ }d}
-2R_{acd\lambda}
R_{b}^{\mbox{ }cd\lambda}
\ee
the order 2 Lovelock tensor.
In what follows we will explicitely set $\alpha_0=-2\Lambda$, $\alpha_1=1$ (expect in section where we will set $\alpha_1=\zeta$) and $\alpha_2=\hat{\alpha}$ for the Gauss-Bonnet coupling.

\section{Exact solutions}

\subsection{A staticity theorem}

We argued in the previous section that Lovelock theory is the natural generalisation of GR in higher dimensions. 
Which of the classical properties of GR remain true when we switch 
on the extra Lovelock densities. One main difference is that for $k \geq 2$, ie once we allow for the Gauss-Bonnet term, (\ref{chahat}), the Weyl tensor appears in the field equations. Therefore Ricci flat solutions are no longer solutions of Lovelock gravity if they are not conformally flat. This means in particular that construction methods quite common in higher dimensional GR are not going to give generically solutions for Lovelock theory. For example, it is not clear on how one can obtain a simple solution such as the black string \cite{charoy}, for Lovelock theory \cite{chakastor} and only a linear correction is known \cite{chakob}. On the other hand we can question the status of classical GR theorems in Lovelock theory such as that of Birkhoff's theorem. Is a version of this theorem still true in Lovelock theory?

Consider the  $D$-dimensional EGB action,
\ba
\label{chaaction21}
S_{(2)}&=&{M^{D-2}\over 2}\int d^D x\, \sqrt{-g}\left[R-2\Lambda +\hat{\alpha} \hat{G} \right] \nonumber\\
\ea
where $M$ is the fundamental mass scale of the D-dimensional theory, and $\hat{G}$ is the Gauss-Bonnet density (\ref{chahat}). We note by 
$2\Lambda=-k^2(D-1)(D-2)$  the bulk bare negative cosmological constant ($2\Lambda=a^2(D-1)(D-2)$ is the positive cosmological constant),
$\hat{\alpha}$ is the Gauss-Bonnet coupling which has dimensions of length squared. We will set $\alpha=(D-3)(D-4)\hat{\alpha}$ to somewhat simplify notation. This action is that of Lovelock theory in component language for $D=5$ or $6$ dimensions. Let us for simplicity and without 
loss of generality \cite{chazeglov} stick to $D=5$ for the rest of this section in order to expose the staticity theorem. 
We will comment on the general result also involving an electric/magnetic field at the beginning of the next section.

Suppose spacetime has constant
3-dimensional spatial curvature. This is the basic hypothesis we will make here.
This is a slight generalisation from the common assumption of spherical symmetry in Birkhoff's theorem. 
A general metric anzatz for the given symmetry is,
\be
\label{chametric1}
ds^2=e^{2\nu(t,z)}B(t,z)^{-\frac{D-3}{D-2}}(-dt^2+dz^2)+B(t,z)^{\frac{2}{D-2}}
\left({d\chi^2\over
1-\kappa \chi^2}+\chi^2d\Omega_{D-3}^2\right)
\ee
where $B(t,z)$ and $\nu(t,z)$ 
are the unknown component fields of the metric and $\kappa=0, 
\pm 1$ is
the normalised curvature of the 3-dimensional homogeneous and 
isotropic surfaces. 
We choose to use the conformal gauge in order to take full advantage 
of the
2-dimensional conformal transformations in the $t-z$ plane. 
The field
equations we are seeking to solve are found by varying the above action
(\ref{chaaction21}) with respect to the background metric or using (\ref{ll}) from the previous section and read 
\ba
\label{chafield2}
\E_{ab}&=&G_{ab}+\Lambda g_{ab}-
\alpha H_{ab}=0
\ea
where $H_{ab}$ is given by (\ref{chalov23}).

It is rather useful to review and compare 
the equivalent system  \cite{chapeter}, \cite{charuth0} in Einstein gravity for $\alpha=0$ where,
$$
R_{ab}=-{(D-3)\Lambda\over (D-2)}g_{ab}.
$$
Pass to light-cone coordinates, 
\be
\label{lightcone}
u={{t-z}\over 2},\qquad v={{t+z}\over 2}.
\ee
and take the
combination $R_{tt}+R_{zz}\pm 2R_{tz}=0$, one obtains the equations which read
\ba
\label{prc}
B_{,uu}-2\nu_{,u} B_{,u}&=&0,\\
B_{,vv}-2\nu_{,v} B_{,v}&=&0.
\ea
Note then that these
are ordinary differential equations with respect to $u$ and $v$.
They are directly integrable giving
\be
\label{chapeter}
B=B(U+V)\qquad e^{2\nu}={B'}{U'}{V'}
\ee
where $U=U(u)$ and $V=V(v)$ are arbitrary functions of $u$ and $v$, 
and a prime stands for the total derivative of the function with 
respect to its unique variable. We refer to (\ref{prc}) as the integrability conditions.
Using a 2 dimensional conformal transformation which is a symmetry of (\ref{chametric1}),
$$
U={{\tilde{z}-\tilde{t}}\over 2},\qquad  V={{\tilde{z}+\tilde{t}}\over 2}$$
gives that the solution is locally static $B=B(\tilde{z})$ and
the equivalent of a generalised Birkhoff's theorem is therefore true. Starting from a general time and
space dependant metric, spacetime has been shown to be locally static or
equivalently that there exists a locally timelike Killing vector field (here
$\partial\over \partial\tilde{t}$). By use of the
remaining field equations we can then find the form 
of $B$, leading after
coordinate transformation to a static black hole solution of horizon curvature $\kappa$.

Let us now take $\alpha\neq 0$. 
In analogy to the previous case 
let us take the combination, $\E_{tt}+\E_{zz}
\pm 2\E_{tz}=0$. On passing to light cone coordinates
(\ref{lightcone}) we get after some manipulations
\ba
\label{integrability}
\left(9B^{4/3}e^{2\nu}+18\alpha \kappa B^{2/3} e^{2\nu}
+2\alpha B_{,u}B_{,v}\right)(B_{,uu}-2\nu_{,u} B_{,u})&=&0\nonumber\\
\left(9B^{4/3}e^{2\nu}+18\alpha \kappa B^{2/3} e^{2\nu}+
2\alpha B_{,u}B_{,v}\right)(B_{,vv}-2\nu_{,v} B_{,v})&=&0
\ea
Note how the Gauss-Bonnet terms factorise nicely leaving the
integrability equations (\ref{prc}) we had in the absence of $\alpha$.

The degenerate case where either 
$B_{,u}=0$ or $B_{,v}=0$  
corresponds to flat solutions. 
For $B_{,u}\neq 0$ 
and $B_{,v}\neq 0$ the situation is clear: 
either we have static solutions and the staticity 
theorem holds as in the case above or we will have 
\be
\label{class1}
e^{2\nu}={2\alpha(B_{,z}^2-B_{,t}^2)\over {9B^{2/3}(B^{2/3}
+4\alpha\kappa)}}
\ee

Let us briefly examine the latter case, that we will call Class I 
solution \cite{chawheeler}, \cite{chafax1}. 
The two remaining field equations $\E_{\chi\chi}=0$ and 
$\E_{tt}-\E_{zz}=0$ are solvable iff we have the simple algebraic
relation,
\be
\label{fine}
4\alpha k^2=1
\ee
This is quite remarkable: if the coupling constants obey (\ref{fine}) 
then the $B$ field is an {\it arbitrary} function of space and
time; in other words the field equations do not determine the metric functions. 
Therefore strictly speaking Birkhoff's theorem does not hold for
non zero cosmological constant. {\footnote{Note however that for a
non-zero charge $Q$ and spherical symmetry ($\kappa=1$) Birkhoff's theorem is
always true as was first shown by Wiltshire \cite{chawiltshire0} }} 
Setting $B(t,z)=R^3(t,z)$ the class I metric reads,
\be
\label{CL12}
ds^2={R_{,z}^2-R_{,t}^2\over
{\kappa+{R^2\over {2\alpha}}}}(-dt^2+dz^2)
+R^2\left({d\chi^2\over
1-\kappa \chi^2}+\chi^2d\Omega_{II}^2\right)
\ee
This solution has generically a curvature singularity for
$R_{,z}=\pm R_{,t}$.
The Class I static solutions are given by,
\be
\label{CL1static}
ds^2=-{A(R)^2\over
{\kappa+{R^2\over {2\alpha}}}}dt^2+{dR^2\over\kappa+{R^2\over {2\alpha}}}
+R^2\left({d\chi^2\over
1-\kappa \chi^2}+\chi^2d\Omega_{II}^2\right)
\ee
with $A=A(R)$ now an arbitrary function of $R$. 

In order to obtain $t$ and $z$ dependent solutions it suffices to 
take
the functional $R$ to be a non-harmonic function. 
Take for instance $R=exp(f(t)+g(z))$,
with $f$ and $g$ arbitrary functions of a timelike and spacelike
coordinate respectively. Let us also assume $\kappa=0$ 
for simplicity,
the Class I metric in proper time reads,
\be
\label{CLtz}
ds^2=-d\tau^2+
{2\alpha dg^2\over {1+2\alpha f_{,\tau}^2}}+e^{2(f+g)}\left({d\chi^2\over
1-\kappa \chi^2}+\chi^2d\Omega_{II}^2\right).
\ee
Note here again that $f$ is an arbitrary function of time and $g$ an arbitrary function of space. The fine tuning 
relation between $\alpha$ and $k$ actually corresponds to a case of enhanced symmetry often refered to as Chern-Simons gravity (for odd dimensional 
spacetimes \cite{chazan}). 

On the
other hand if (\ref{fine}) does not hold then Birkhoff's theorem 
remains
true in the presence of the Gauss-Bonnet terms i.e. {\it the general 
solution assuming the presence of a cosmological constant in the bulk
and 3 dimensional constant curvature surfaces is static if and only
if (\ref{fine}) is not satisfied}. 
In this case the remaining two equations give the same ordinary
differential equation for $B(U+V)$ which after one integration reads,
\be
\label{B}
{B'}+9B^{2/3}(k^2 B^{2/3}+\kappa)+
9\alpha\left({B'\over {9B^{2/3}}}+\kappa\right)^2=9\mu
\ee
where $\mu$ is an arbitrary integration constant. Then by making $B$ 
the
spatial coordinate and setting $B^{1/3}=r$ we get the solution 
discovered and discussed in
detail by Boulware-Deser \cite{chaboul} 
($\kappa=1$) and Cai \cite{chacai} ($\kappa=0,-1$),\footnote{We have kept
the same label as in
(\ref{chametric1}) for the rescaled time coordinate .}
\be
\label{BH}
ds^2=-V(r) dt^2+{dr^2\over V(r)}+r^2\left({d\chi^2\over
1-\kappa \chi^2}+\chi^2d\Omega_{II}^2\right)
\ee
where $V(r)=\kappa+{r^2\over
{2\alpha}}[1\pm\sqrt{1-4\alpha k^2+4{\alpha \mu\over r^4}}]$, and
$\mu$ is an integration parameter related to the gravitational mass. 
The maximally symmetric solutions are obtained by setting $\mu=0$.
We will analyse in detail these solutions in the following section.

To close off 
notice how (\ref{fine}) is a particular 'end' point for (\ref{BH})
since the maximally symmetric solution is defined only for $1\geq 4\alpha
k^2$ (for $\alpha<0$ there is no such restriction). We can deduce in all
generality that
for $1\geq 4\alpha k^2$ there is a unique static solution (\ref{BH}).
When (\ref{fine}) is satisfied 
and $\mu=0$, 
the two branches coincide (we have an infinity of solutions) and 
$V=\kappa\pm\sqrt{\mu/\alpha}+{r^2\over {\alpha}}$, 
is then a particular  Class I solution (\ref{fine}) 
very similar to the BTZ 3-dimensional black holes. For $1\leq 4\alpha k^2$ no solutions exist.
We will come back to the 6 dimensional version of these black holes in a moment. Before doing so let's see briefly how they generalise for the full Lovelock theory.

\subsection{Lovelock black holes}

The staticity theorem we evoked in the previous subsection is generalised without major difficulty for the 
general Lovelock theory in arbitrary dimension and in the presence of an Abelian gauge field \cite{chazeglov}. 
That is for the theory involving a Lovelock action (\ref{lla}) with a Maxwell gauge field,
\be
\label{chalov22}
S_{D}^{EM}=S_D-\frac{1}{4}\int d^Dx \sqrt{-g} F_{ab}F^{ab}
\ee
we obtain from the field equations that apart from the pathological cases of Class 1 there is a unique solution given by \cite{chams}
\be
\label{chalov}
ds^2=-V(r) dt^2+\frac{dr^2}{V(r)}+r^2 (\frac{d\chi^2}{1-\kappa x^2}+\chi^2 d\Omega^2_{D-3})
\ee
where the electric field strength is $F=\frac{q^2}{4\pi r^{2(D-2)}}\; dt\wedge dr$.
The metric potential reads $V=\kappa-r^2 f(r)$ where $f$ is a solution of the $k$-th order algebraic equation,
\be
\label{chapoly}
P(f)=\sum_{l=0}^{k} \hat{\alpha_l}\, f^l=\frac{\mu}{r^{D-1}}-\frac{Q^2}{r^{2D-4}}
\ee
where $k=\left[\frac{D-1}{2}\right]$ is the order of the Lovelock theory. It is clear that the higher the dimension of spacetime, the more the terms in the Lovelock action, and hence the higher the order of the equation (\ref{chapoly}). The $k$ possible roots of the polynomial  $P(f)$ for $\kappa=1$ actually give us the maximally symmetric vacua of the theory \cite{chams}. Note that even in the absence of a bare cosmological constant these vacua can be flat or of positive or negative curvature and their magnitude depends on the normalised Lovelock coupling constants,
\ba
\hat{\alpha}_0&=&\frac{\alpha_0}{\alpha_1}\frac{1}{(D-1)(D-2)},\qquad \hat{\alpha}_1=1 \nonumber,\\
\hat{\alpha}_l&=&\frac{\alpha_l}{\alpha_1}\prod_{n=3}^{2l} (D-n), \texttt{for} \quad l>1
\ea
Positive roots correspond to de Sitter vacua whereas negative roots to anti de Sitter vacua. It is interesting to note that in the presence of a bare cosmological constant $\alpha_0\neq 0$ maximally symmetric vacua may not exist at all.  For zero bare cosmological constant however the flat vacuum is always solution. 
Solutions of (\ref{chapoly}) have complex horizon structures and have been analysed by Myers and Simon \cite{chams} for $Q=0$. 

\subsection{Einstein-Gauss-Bonnet black holes}

For simplicity, let us now truncate Lovelock theory at $k=2$ 
ie neglect higher order terms other than the Gauss-Bonnet invariant. We will follow for most of this section the analysis
of \cite{chams}.  
We therefore consider the action (\ref{chalov22}) for up to $k=2$,
\be
\label{chachouk}
S_2^{EM}=S_{(2)}-\frac{1}{4}\int d^Dx \sqrt{-g} F_{ab}F^{ab}
\ee 
Take a $(D-2)$ dimensional space of maximal symmetry and therefore of constant curvature parametrised by $\kappa=0,-1,1$,
\be
h^{S}=h^{S}_{\mu\nu}dx^\mu dx^\nu = \frac{d\chi^2}{1-\kappa \chi^2}+\chi^2 d\Omega_{(D-3)}^2
\ee
These are the maximal symmetry spaces we considered in  the previous section and express the constant curvature 
geometry, normalised to $\kappa$, of the horizon surface. 
If on the other hand we perform a careful Wick rotation to $h^S$ we can construct $h^{L}$, which is of Lorentzian signature, and gives   spacetimes
 which are $(D-2)$-dimensional sections of Minkowski, adS and dS. Therefore, the staticity theorem of the previous section tells 
us that any $D$ dimensional spacetime metric admitting $D-2$ dimensional maximal sub-spaces $h^S$ (or sub-spacetimes $h^L$) which 
is solution of the field equations emanating from (\ref{chaaction21}) is locally isometric to,
\be
\label{chabh}
ds^2=-V(r) dt^2+\frac{dr^2}{V(r)}+r^2 h^S_{\mu\nu} dx^\mu dx^\nu
\ee
with electric field strength
\be\label{chafs}
F=\frac{q^2}{4\pi r^{2(D-2)}}\,  dt\wedge dr
\ee
or, modulo a double Wick rotation, to 
\be
\label{chabh1}
ds^2=V(r) d\theta^2+\frac{dr^2}{V(r)}+r^2 h^L_{\mu\nu} dx^\mu dx^\nu
\ee
with magnetic field strength
\be\label{chafs1}
F=\frac{p^2}{4\pi r^{2(D-2)}}\,  d\theta\wedge dr
\ee
In the latter case the theorem gives a local axial Killing vector $\partial_\theta$ and 
concerns locally axially symmetric solutions.  This case has been explicitely treated for the case of general relativity $k=1$ in \cite{chagrepad}. 
We will come back to this 
simple yet powerful result to give the maximally symmetric cosmic string metrics in $D=4$ 
and maximally symmetric braneworlds of codimension 2 in $D=6$.

Since we want to search for black hole criteria we concentrate on the static case for the rest of this subsection. 
The potential reads (as can be easily verified from (\ref{chapoly})),
\be
\label{chapot}
V(r)=\kappa+\frac{r^2}{2\alpha}\left[1+\epsilon \sqrt{1+4\alpha(a^2+\frac{\mu}{r^{D-1}}-\frac{Q^2}{r^{2(D-2)}})}\right]
\ee
with integration constants,
\be
\label{chapar2}
Q^2=\frac{q^2}{2\pi (D-2)(D-3)}, \qquad \mu=\frac{16\pi G M}{(D-2) \Sigma_{\kappa}}
\ee
where $q$ is the charge, $M$ is the AD or ADM mass of the solution and $\Sigma_{\kappa}$ is the horizon area.

First notable fact is the the ambiguity of the vacuum which is parametrised by $\epsilon=\pm1$ and gives rise to 2 branches of solutions. Indeed setting $\mu=Q^2=0$, $\kappa=1$ gives us the possible vacua of the theory (\ref{chaaction21}). 
In fact we note that if we set the bare cosmological constant to be zero $a=0$ we note that we do not only obtain the flat vacuum. For $\epsilon=1$, we actually asymptote anti de Sitter space for $\alpha>0$ and dS space for $\alpha<0$. We will refer to this vacuum, as the Gauss-Bonnet branch. Indeed the effective cosmological constant is $\Lambda_{eff}=\frac{(D-1)(D-2)}{\alpha}$ and hence $\alpha$ plays this role without a bare cosmological constant term in the 
action. Unlike what was first argued in \cite{chaboul} this branch is not, at least, classically unstable \cite{chadestek} in the sense that 
one needs to add positive energy to the system for it to roll off to a positive mass black hole solution. We will here caution the reader 
that this branch is still dangerous for stability and a further careful analysis is still needed to answer the question of the physical 
relavance of this branch \cite{chahoolio}. It is an intriguing fact however, that one can have an effective cosmological constant from higher order curvature 
terms as used in \cite{chatolley}. Any solution of this branch will not have an Einstein theory limit, although there maybe a relevant de 
Sitter or anti de Sitter Einstein type solution mimicking (\ref{chabh}) as we will see in a moment.  
The $\epsilon=-1$ branch gives the usual Minkowski vacuum in the absence of a bare cosmological constant $\Lambda$ and for large $r$ the 
solutions resemble the asymptotically Einstein black holes with the relevant mass and charge parameters. This branch therefore is perturbatively connected to Einstein theory and we can consider the action (\ref{chachouk}) as an effective action including a higher order correction as in actions for closed strings \cite{chagross}. For the Gauss-Bonnet branch $\epsilon=1$ notice that this is no longer the case and small $\alpha$ yields an enormous effective cosmological constant. Hence it is in this case an erronius statement to consider (\ref{chachouk}) as an effective action and one needs exact solutions rather than perturbative ones. This fact makes this branch either totally irrelevant or far more interesting than the Einstein branch since this is where we can expect novel effects. Given therefore the fact that the $\epsilon=-1$ case closely follows known solutions we will check out mostly the Gauss-Bonnet branch for novel effects.

The ``Chern-Simons'' limit is obtained for $4\alpha k^2=-4\alpha a^2=1$ (for $\alpha$
 positive and negative respectively). The two branches then become one and at this limit the combined Gauss-Bonnet and cosmological 
constant are of the same order as the Einstein term on the right hand side 
(whose coupling we have normalised to 1). For zero charge the potential simplifies to the simple function
\be
V(r)=\kappa+\frac{r^2}{2\alpha}-\frac{m}{\sqrt{r^{(D-5)}}}
\ee
where $m=-\sqrt{\frac{\mu}{\alpha}}$ is the mass integration constant. For $D=5$ the potential is quite similar to the BTZ black holes of 3 dimensions since the mass plays the same role as curvature of the horizon 
$\kappa$. Furthermore note that we have a weaker gravitational force than in the Einstein case. 
Indeed set $D=6$ and a planar black hole for $\alpha>0$ exists with $r_h>(2m\alpha)^{2/5}$ whereas 
the Einstein horizon for the same mass $m$ is its square root. Hence for the same mass the Chern-Simons black hole is of squared 
horizon radius compared to the Einstein one.
Seemingly the closer we approach the Chern-Simons branch the milder  is the curvature singularity  in the bulk spacetime an interesting 
fact to keep in mind.

So much for the branches of solutions.
Let us now check out the curvature singularities and horizons.
Generically, there are two possible singularities in the curvature tensor, the usual $r=0$, but also a branch singularity at the maximal possible zero of the square root, say $r=r_1$.  We have that $r=r_1$ is solution of 
\be
\label{chasing}
r^{2(D-2)}(1+4\alpha a^2)+4\alpha(\mu r^{D-3}-Q^2)=0
\ee
and whenever  $r_1>0$ this is the singular end of spacetime (\ref{chabh}). We have a black hole solution if and only if there exists $r$ spacelike with $r=r_h$ such that $V(r_h)=0$ and $r_h>r_{min}$, where $r_{min}=max({0,r_1})$. Indeed the usual Kruskal extension 
\be 
dv_{\pm}=dt\pm \frac{dr}{V(r)}
\ee
gives that $(v_+,r)$ and $(v_-,r)$ constitute a regular chart across the future and past horizons of (\ref{chabh}) as in GR.
It is straightforward to show the following criterion : $r=r_h$  is an event horizon for $r>r_h$ spacelike iff,
\begin{itemize}
\item{$r_h>r_{min}$}
\item{$\epsilon(2\alpha \kappa+r_h^2)\leq 0$}
\item{$r_h$ is root{\footnote{For $r$ spacelike we need $p(x)>0$ away from the horizon for $\alpha\geq 0$ and the contrary for $\alpha<0$ }} of $p_\alpha(x)=(-a^2 x^{2(D-2)}+\kappa x^{2(D-3)} +\alpha \kappa^2 x^{2(D-4)}-\mu x^{D-3} +Q^2) sign(\alpha)$}
\end{itemize}
Notice therefore that for $\epsilon=1$ we have $r_h^2\leq -2\alpha \kappa$ ie the event horizon is bounded from above. This immediatelly yields that for $\kappa=0$ there are no black hole solutions in this branch! Secondly, note that $p_{\alpha=0}$ is the usual polynomial for $\Lambda$-RN black hole of Einstein theory in $D$ dimensions. In particular, $\alpha$ couples only to the horizon curvature. Hence for $\epsilon=-1$ and $\kappa=0$ (the Einstein branch) the horizon positions are the same as for the GR planar black holes. 

Let us now focus on some particular solutions for the Gauss-Bonnet branch (again we advise the interested reader to see the nice analysis of \cite{chams}). Thus take $\epsilon=1$, $a=0$, $\kappa=1$. We have $r_h^2\leq -2\alpha$, hence take $|\alpha|=-\alpha>0$. We also take $\mu<0$  in order for $(\ref{chasing})$ to be strictly positive and to have a correct definition of mass. Indeed the solution asymptotes a de Sitter Schwarzschild solution with positive AD mass for negative $\mu$. Now given that $p_\alpha(x)=-x^3(x^3-x|\alpha|-\mu)$ we find $x_{min}=\sqrt{\frac{|\alpha|}{\sqrt{3}}}$ and we have two event horizons as long as $-\frac{2}{3\sqrt{3}}|\alpha|^{3/2}\leq \mu < 0$. The bigger the mass of the black hole the further we are allowed to stretch the $\alpha$ parameter of the action (see figure). The solution has the same structure as Schwarzschild de Sitter with an event horizon and a de Sitter horizon. If we switch on $Q^2$ we can get a 3 horizon structure  similar to the de-Sitter RN black hole. When the two horizons come together at $x_{min}$ we are in the extremal case of 
the Nariai solution.

Consider now a hyperbolic horizon, in other words $\kappa=-1$. Take $Q=a=0$ and $\epsilon=1$. We have $r_1^5=-4\alpha \mu$ and $r_h^2\geq -2\alpha$. Therefore for $\alpha>0$ and $\mu>0$ we have trivially the first two conditions verified. Furthermore,  $r_h$ is root of $-x^3+\alpha x -\mu$ which is exactly the same polynomial as above for $\alpha\rightarrow -\alpha$ and $\mu \rightarrow -\mu$. Therefore in this case we will have a hyperbolic black hole with 2 event horizons.

	\begin{center}
	\includegraphics[bb=0 0 427 192]{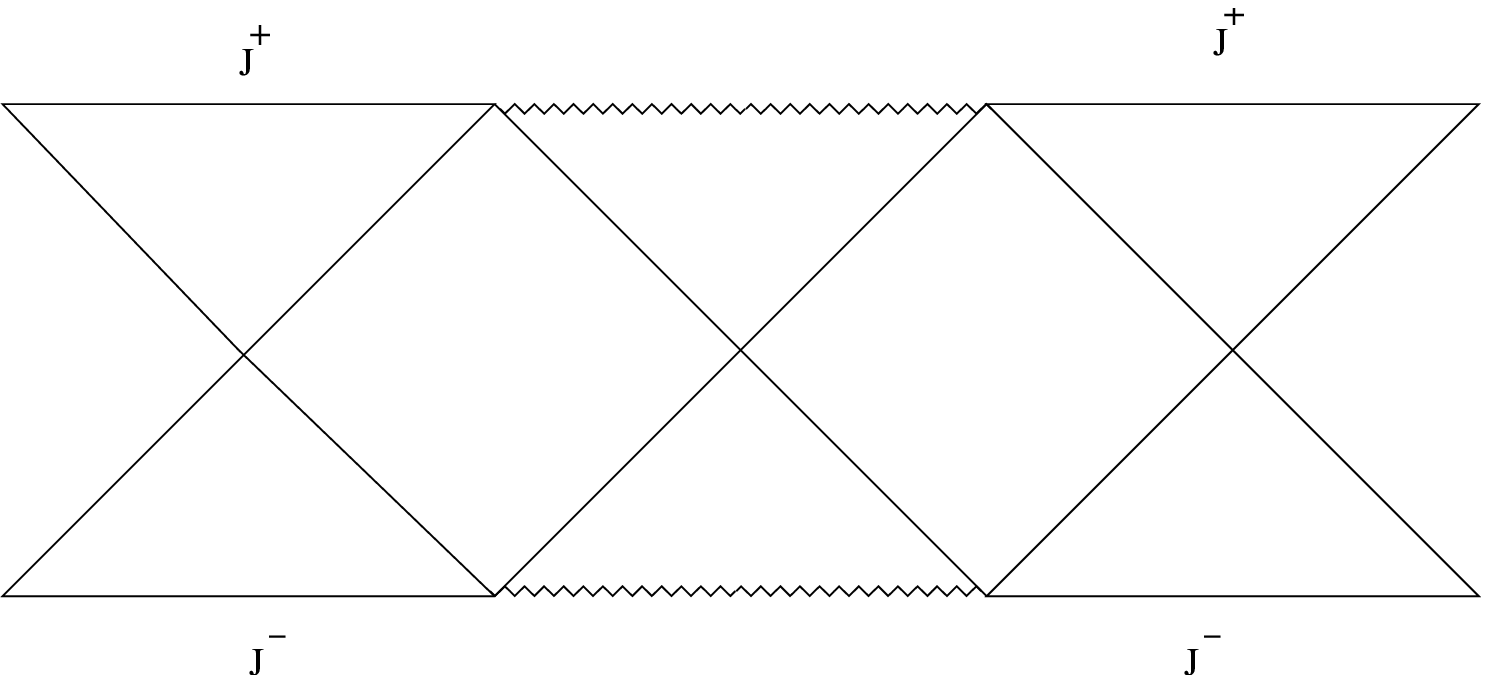}
	\end{center}

\subsection{Thermodynamics and a geometrical explanation of the horizon area formula}

Our aim here is to evaluate the entropy of the black hole solutions descussed in the previous section. We therefore start by evaluating the mass and temperature.
The mass of the black hole can be easily  expressed in terms 
of the horizon radius, $r_h$ by,
\be
\label{chamass}
M=\frac{(D-2)\Sigma_\kappa}{16\pi G}(\kappa-r_h^2 a^2+\frac{\alpha \kappa^2}{r_h^2}+\frac{Q^2}{\alpha r})r_h^{D-3}
\ee
In order to calculate the temperature of the black hole we follow the standard prescription. In summary: we start by Wick rotating the time direction $t\rightarrow i\theta$. The resulting curved manifold of Riemannian signature  has an axial Killing vector $\partial_\theta$ at its origin situated at $r=r_h$  as we will see in a moment. We then impose periodicity, say $\Pi$, in order for the angular coordinate to be well defined. As a result the Euclidean quantum field propagator, with the imposed periodic boundary conditions, describes a canonical ensemble of states in thermal equilibrium at a heat bath of temperature $T=\Pi^{-1}$ \cite{chagibb}.

Indeed consider $t \rightarrow i \theta$ of (\ref{chabh}). We have
\be
\label{chains}
ds^2=V(r) d\theta^2+\frac{dr^2}{V(r)}+r^2 h^S_{\mu\nu} dx^\mu dx^\nu
\ee
\be
\label{chapot1}
V(r)=\kappa=\frac{r^2}{2\alpha}\left[1+\epsilon \sqrt{1+4\alpha(a^2+\frac{\mu}{r^{D-1}}+\frac{P^2}{r^{2(D-2)}})}\right]
\ee
with (magnetic) field strength
\be\label{chafs2}
F=\frac{p^2}{4\pi r^{2(D-2)}} d\theta\wedge dr
\ee
and the metric is then of Euclidean signature for $r>r_h$ and $p=iq$  ($P=iQ$) is the magnetic charge. Let us put the charge equal to zero for simplicity. The potential $V(r)$ admits at least one and at most  2 zeros for $0<r_-<r_+$ where $r=r_+$ can be taken to be infinity for the single horizon case. Taking $x^\mu=constant$ and expanding around the zeros of the potential V  we get,
\be
\label{chacyl}
ds^2\sim \left(\quarter V_{r_\pm}'^2\right) \rho^2_{\pm}d\theta^2+ d\rho_\pm^2
\ee
with radial isotropic (or cylindrical) coordinate $\rho_\pm=\sqrt{\frac{2(r-r_\pm)}{V'_{r_\pm}}}$. Hence the periodicity condition reads,
\be
\label{chareg}
\theta \sim \theta+\frac{4\pi}{V'_{r_-}}\sim \theta +\frac{4\pi}{V'_{r_+}}
\ee

Generically we will have conical singularitites. In fact the only "finite" regular solution exists when, $V_{r_-}'=V_{r_+}'$ ie  for a $T=0$ temperature black hole (Nariai limit) 
and it corresponds to a regular gravitational  instanton. In fact, one has to consider a simple coordinate transformation 
to show that space is of non zero volume and corresponds to $S^4\times S^2$ check out \cite{chaperry}, \cite{chaantonis}. When we add charge it is a question of algebra to show that we can obtain warmer, $T\neq 0$ instantons{\footnote{I thank Renaud Parentani for pointing this out}}. Gravitational instantons 
describe the extremal spacelike path between two non-causaly 
connected regions of gravitational solutions \cite{chahawhor}, in other words their quantum tunneling. In the case here the instanton describes decay of  the vacuum 
into a pair of accelerated black holes \cite{chaantonis} as in usual GR (for instanton solutions using boundaries see \cite{chagrav}). 
The decay rate of course would have to be calculated here from scratch since it corresponds to a novel theory \cite{chaantonis}. If it is enhanced it corresponds to a quantum instability of the vacuum. 
When $r_+=\infty$ space rounds up smoothly and regularity is only imposed at $r=r_-$. 

After this brief parenthesis let us now calculate the temperature of the black holes (\ref{chabh}). The period is $\Pi=\frac{4\pi}{V'_{r_\pm}}$. It is easy to show the formula,
\be
V'_{r_\pm}=\frac{1}{r_\pm}\left[ \frac{\mu(D-1)}{r_\pm^{D-5}(2\alpha \kappa+r_\pm^2)}-2\kappa \right]
\ee
The heat bath described by the thermal propagator is of temperature $T=\Pi^{-1}$. Therefore the temperature is found to be (using (\ref{chamass})),
\be
\label{chatemperature}
T=\frac{(D-1)k^2 r_\pm^{D-1}+(D-3)\kappa  r_\pm^{D-3}+(D-5)\alpha\kappa^2 r_\pm^{D-5}}{4\pi^2 r_\pm(r_\pm^2+2\alpha \kappa)}
\ee
 Notice that the Gauss-Bonnet term $\alpha$ couples only to the horizon curvature and the temperature is the same
 for $\epsilon=\pm 1$. Also note that  a Lovelock planar black hole
 will have same temperature as an Einstein one \cite{chacai}. For the entropy we use here the standard 
recipe, $dM=TdS$ (see also  \cite{chakofi3} for a direct calculation yielding the  same result) . Then in turn,
\ba
S=\int T^{-1}dM&=&\int_{r_{min}}^{r_h} T^{-1} \frac{\partial M}{\partial \bar{r}_h} d\bar{r}_h=\nonumber\\
&=&\frac{r_h^{D-2} \Sigma_\kappa}{4G}(1+\frac{2\alpha \kappa(D-2)}{(D-4) r_h^2})
\ea
In order to evade erronius conclusions it is important to note that $r_{min}=max(r_1,0)$ \cite{chaclunan}, since in Lovelock black holes we can hit a singularity before reaching $r=0$.
Furthermore we see that the entropy is not equal to the area of the black hole horizon; we pick up a correction from the induced curvature of the horizon surface. One can understand this fact geometrically using the general formalism of Iyer and Wald, \cite{chaiyer}. The entropy calculation in this case \cite{chaclunan} gives the  result,
\be
S=\frac{1}{4G}\int_{r_{min}}^{r_h} dx^{D-2}\; \sqrt{\tilde{h}_{\mu\nu}}(1+2\alpha \tilde{R}^{ind})
\ee
where the tilded terms correspond to geometrical quantities of the horizon surface, $(r,\theta)=$constant. The result 
is exactly the same as the effective action obtained for a codimension 2 matching condition (\ref{melina}) as we will see in the forthcoming section{\footnote{Here the extrinsic curvature quantities appearing in (\ref{melina}) cancel because the horizon surface is maximally symmetric}}.. This is not too surprising with hindsight: In both cases we impose similar regularity conditions which have to do with the temperature and hence the periodicity of the manifold. In one case conical singularities are accounted for by the presence of branes of given tension whereas for black holes this precicely gives the temperature of the horizon. The leading Lovelock correction appearing as an induced curvature term is therefore nothing but the extended Euler density of the horizon surface. On the other hand the Einstein term yields the tension of the horizon. Therefore the failure to obey in this case a horizon-area formula is most natural and a simple geometric consequence. The entropy of the Lovelock black holes follows the matching condition formula. For example if we were to take a $D=8$ dimensional Lovelock black hole we would get in addition the 6 dimensional Gauss-Bonnet correction of the horizon surface in the entropy formula namely,
\be
S=\frac{1}{4G}\int dx^{D-2} \sqrt{\tilde{h}_{\mu\nu}}(1+2\alpha \tilde{R}^{ind}+\alpha_2 \hat{G}^{ind})
\ee
We will come back to these formulae in the next section.

\subsection{The Lovelock solitons and the maximally symmetric cosmic strings}

So much for the Euclidean version of (\ref{chabh}) and the thermodynamics (for details check \cite{chaclunan}). We now Wick rotate the $D-2$ horizon sections $h^S_{\mu\nu}$ to the Lorentzian maximally symmetric sections $h^L_{\mu\nu}$ to construct soliton solutions (\ref{chabh1}) (the same obviously holds for the Lovelock case (\ref{chalov})).  Since the soliton is of axial symmetry we have $r_-\leq r\leq r_+$ where $r_\pm$ are the possible zero's of V. One can always locally go to the cylindrical anzatz (\ref{chacyl}) with $\rho_\pm$ the local cylindrical coordinate. 
The soliton is of infinite proper distance $\rho$ when $r_+\rightarrow \infty$. A very nice and simple interpretation of these solutions and a simple application of the statiticity theorem resides in $D=4$ general relativity.  In this case the Lorentzian 2-dimensional sections in (\ref{chabh1}) have precicely the geometry of a maximally symmetric cosmic string, of internal constant curvature given by $\kappa$, situated at $r=r_\pm$ once we allow for some deficit angle in $\theta$. These are the only solutions describing de-Sitter, flat or anti de-Sitter  cosmic strings in a cosmological constant and (possibly)  charged background. We will not analyse these here but as an example it is  easy to note that in the presence of a bare negative cosmological constant the straight cosmic string bends the ambient spacetime (this was first noted by Linet in \cite{chalinet}). This is in complete contrast to the flat string which is embedded in a locally flat spacetime. This is because  the resulting geometry is of non-trivial Weyl tensor,
\be
\label{linet}
ds^2=(k^2r^2-\frac{\mu}{r})\beta^2 d\theta^2+\frac{dr^2}{(k^2r^2-\frac{\mu}{r})}+r^2 (-dt^2+dz^2)
\ee
precicely denoted here by the presence of a non-trivial $\mu$ parameter.
In (\ref{linet}) $(t,z)$ are the string coordinates, the angular deficit is given by $\delta=2\pi(1-\beta)$ and the string is of linear tension $T=\frac{\delta}{8\pi G_4}$.
 In other words a straight cosmic string mathematically at least corresponds to  a double Wick 
rotated black hole solution. Another way to 
understand the above  result is that in pure adS spacetime we cannot slice the geometry in a cylindrically symmetric anzatz. It is a simple exercise to classify the solutions, the case of de-Sitter also presenting some unique characteristics due to the compacteness of its spatial sections (cosmic strings will appear in pairs in this case).
Furthermore,  for $\kappa=0$, $D$ arbitrary and negative cosmological constant we get the Lovelock ads soliton analysed in \cite{chamyho} for the case of Einstein gravity. In a forthcoming section we will see that in $D=6$ dismensions the metrics (\ref{chabh1}) will describe (with the help of suitable matching conditions) the generic bulk geometry of maximally symmetric 4 dimensional braneworlds \cite{chaantonis}.

\section{Matching conditions for distributional sources}

Most recent applications of Lovelock theory concern braneworld physics. In the braneworld picture our 4 dimensional universe 
is part of a higher dimensional manifold and whereas matter is strictly confined on the braneworld,
gravity propagates, according to the equivelance principle, in all dimensions. We therefore 
want to describe, the motion or the evolution, of a self-gravitating submanifold which for simplicity we take to be infinitesimaly thin. In other words, we suppose that matter is confined on the braneworld via a Dirac distribution of dimension equal to the codimension $N=D-4$ of the braneworld. Given that Lovelock theory is the general metric theory with second order field equations we can except some junction conditons quite similar to those of Einstein theory{\footnote{Fourth-order field equations would mean that distributional constraints would be imposed on the continuity of certain third order directional derivatives which would mean a discontinuous limit to pure Einstein theory.}}. It turns out that in Lovelock theory we can even do better, at least in mathematical terms, and define matching conditions
for higher codimension than that  possible for Einstein theory {\cite{chaparadox}}.

Consider a p-brane $\Sigma$ embedded in
$D=p+1+N$ dimensions and suppose that $\Sigma$   
carries some
 localised 
energy-momentum tensor~{\footnote{Greek letters run through brane coordinates
    while capital Latin from the begining of the alphabet through bulk coordinates}},
\be
\label{chaset}
T_{AB}= \left ( \begin{array}{cc}
S_{\mu \nu} & 0 \\
0&0
\end{array} \right ) \delta_{\Sigma} \, .
\ee
The $N$-dimensional Dirac distribution $\delta_\Sigma$ on $\Sigma$ 
signifies that the brane is of zero thickness. The question we address is:
What are the equations of motion for this self-gravitating p-brane  sourced by
the distributional energy momentum tensor (\ref{chaset})?  
For codimension $N=1$, the answer is given by the well-known 
Israel junction conditions \cite{chaisr} where, if the induced 
metric is continuous,  a discontinuity in the first 
derivatives of the
metric accounts for the Dirac charge in (\ref{chaset}). 
Israel's junction conditions describe adequately a wide variety of GR problems and only when 
we study junction problems of lesser symmetry do we run into shortcomings because we can no longer fullfil the continuity condition (for example we cannot match Kerr to flat spacetime).
If the codimension
is strictly greater than 2 then there are no distributional 
matching conditions in Einstein
gravity. Thus, a finite thickness braneworld is needed in order to 
obtain non-trivial self-gravitating equations
of motion for $\Sigma$ and the resulting equations of motion will generically depend on the regularisation scheme. This is hardly surprising. We know for example, that far away from a gravitational source such as the sun, and by virtue of Birkhoff's theorem, 
we can approximate conveniently its gravitational field using a 3 dimensional Dirac distribution. What we mean by far away is  precicely the Schwarzschild radius of the source in question which is about 3km and which is far smaller than the actual size of the sun. In Einstein theory even when the codimension is equal to 2 (for $D=4$ say), one only knows the self-gravitating field of a straight cosmic string, \ie one induced by a pure
tension matter tensor, which gives an overall conical deficit angle \cite{chaparadox}. Hence even for codimension 2 
distributional matching conditions only give a brute picture of the gravitational field and break down when the braneworld or defect is of lesser symmetry.
The essential point for the discussion  is the codimension of spacetime $N$, ie the number of spacelike or timelike, vectors defined normal to the brane (for lightlike junctions see \cite{chabar}). If the codimension is 1 as it is for the usual junction conditions the $p$-brane is a hypersurface splitting the bulk in two and there is no geometry (other than that of real line) in the normal directions. Once $N>1$ we have non-trivial geometry in the normal sections. 

Lovelock's equations for the distributional energy momentum tensor (\ref{chaset}) are 
\be
\label{chall1}
\sum_{k=0}^{[(D-1)/2]} \alpha_k {\cal E}_{(k) \mu} = -2S_{\mu\nu} \theta^{{\star} \nu} \delta_\Sigma
\, ,
\ee
Our task here involves finding the distributional part of the LHS to be matched to the induced energy-momentum tensor $S_{\mu\nu}$. We will refer to such geometric distributional  terms as those carrying Dirac charge. 
It turns out that not all Lovelock terms can carry a Dirac charge, we already know this from Einstein theory which is Lovelock theory in $D=4$. Indeed we will find the simple inequality, 
 \be
\label{special}
N/2\leq k \leq \left[ {D-1} \over 2 \right] \, ,
\ee
which selects in particular the Lovelock bulk terms $\L_{(k)}$ that can carry
a Dirac charge. Indeed for $D=4$ we have $N/2\leq k\leq 1$ which tells precicely that an Einstein theory, $k=1$
 can carry up to codimension 2 Dirac charge. 
If $D>4$ and we allow for the higher order Lovelock terms we see that we can go to higher codimension. 

Without attempting to give a full proof of this result, given in \cite{chazeg}, we can however understand  it geometrically in the following way: 
seperate the geometry in normal and parallel sections as we commonly do in the Gauss-Codazzi formalism. 
Let $e_\mu$ be the
$p+1$ unit vectors that are everywhere tangent to the brane, $\n_I$ be the
$N$ unit vectors that are everywhere normal to $\Sigma$ with the label $\N$
denoting the radial normal vector. Similarily, split locally the bulk $\theta^A$ into
tangent 1-forms $\theta^\mu$ and normal 1-forms $\theta^I$. One can then
deduce the first and second  fundamental forms of the brane, respectively; 
\ba
h &=& \eta_{\mu\nu} \theta^\mu \otimes \theta^\nu \\
K_{I\mu\nu} &=& g(\nabla_{e_\mu} \n_I, e_\nu) 
\ea 
describing 
the induced metric and extrinsic curvature of $\Sigma$.
The Gauss-Codazzi equations can also be written in form formalism which is extremenly useful in this context. 
Indeed the parallel projection along the brane coordinates gives the Gauss equation,
\be
\label{chagform}
\R^\mu_{{\pp}\nu}\equiv \frac{1}{2}R^\mu{}_{\nu\lambda\rho} \theta^\lambda \wedge \theta^\rho= \Omega^{\mu}_{\pp \nu}- \k^{\mu}_{{\pp}I} \wedge
\k^I_{{\pp}\nu} \, ,
\ee
where 
$$\k^I_{\pp\mu}= K_{\nu \mu}^I  \theta^{\nu} \, ,$$
is the extrinsic curvature 1-form and $\Omega^\mu_{{\pp}\nu}$ is the {\it induced curvature
2-form} of the brane, associated with the induced metric $h_{\mu\nu}$. This geometrical identity relates 
the background curvature with respect to the induced and extrinsic curvature of $\Sigma$.

We look for geometric terms in the
$\mu\nu$-components  on the LHS of (\ref{chall1}) that can carry a Dirac
charge. The situation is inherently different for odd and even codimension. Indeed for hypersurfaces (codimension 1), the distributional charge in Einstein theory is provided by a 
discontinuity of the extrinsic curvature of the metric $K_{\mu\nu}^N$. Thus we relate local geometry of the brane to matter. In even codimension defects, like cosmic strings (codimension 2), the normal sections to the brane have some non-trivial topology that can give rise to distributional terms. 
For the case of the straight cosmic string we have,
\be
\label{chastring}
ds^2=-dt^2+dz^2+d\rho^2+L^2(\rho) d\theta^2+d\rho^2
\ee
and Einstein field equations are solved for $L(\rho)=\rho$ everywhere but at $\rho=0$. We therefore set
$$
L'(\rho)=1, \;\rho\neq 0 \; \mbox{and}  \;L'(\rho)=\beta , \;\rho=0 
$$
The distributional part of  (\ref{chall1}) is then integrated over the normal section and reads
\be
\label{chadist}
\int_0^{2\pi} \int_0^r \mbox{dist}(L'')rdr\, d\theta=-8\pi G_4 \int_\Sigma^\perp \frac{T}{2\pi}\delta^{(2)} \,r dr \, d\theta
\ee
Given that $\mbox{dist}(L'')=[L'] \delta^{(2)}=(1-\beta) \frac{\delta(r)}{r}$, we get the well-known result $2\pi (1-\beta)=-8 \pi G_4T$ relating here topology (rather than geometry) to matter. It is important to note that the cuvature of the normal section is precicely given by $L''$.

In each case we define  locally a normal section,
$\Sigma_\perp$, at each point of the brane, using the congruence of the normal vectors with specific regularity properties. We then
integrate the $\mu\nu$-equations of motion over an arbitrary such section 
$\Sigma_\perp$. Since the Dirac distribution is by definition independent of regularisation ie to the
microscopic features of $\Sigma$, the only terms
that can contribute are those {\it independent} of geometry  deformations over 
$\Sigma_\perp$ as we take the limit of zero thickness. This boils down for codimension 2 to working out the distributional part as in (\ref{chadist}).
Mathematically, the only terms in (\ref{chall1}) having this property are proven to be locally exact forms on 
$\Sigma_\perp$, basically made out of powers of extrinsic curvature or curvature of the normal section as for the example of the cosmic string above{\footnote{In codimension 2 the above statement boils down to the famous Gauss-Bonnet theorem. In higher codimension the relevant forms are precicely the Chern-Simons forms, \cite{chachern}}.

What {\it{multiplies}} these locally exact forms are now equated to the energy-momentum tensor of the brane 
and depend on the parallel geometry (since the normal geometry is integrated out) using essentially Gauss's equation (\ref{chagform}). 
Since higher order Lovelock terms, $k>1$, contain higher powers of the curvature tensor, we can 
expect the appearence of induced curvature and extrinsic curvature terms from (\ref{chagform}). Indeed the relevant terms are sums 
in $0\leq j\leq min([N/2],k-N+[N/2])$ of 
\be
\label{cake}
\sigma_{(N,k,j)\mu}^{\pp}=\left (\bigwedge_{l=1}^{k-N+[N/2]-j} \R^{\lambda_{l} \nu_{l}}_{{\pp}} \right ) \wedge
\left ( \bigwedge_{l=1}^{N-2[N/2]+2j} \k^{\rho_l}_{{\pp} N} \right ) \wedge
\theta^{\star {\pp}}_{\mu \lambda_l \nu_{l} \rho_{l} } \, ,
\ee
which are powers of the projected Riemann curvature two form $\R_{\pp}$ 
and the extrinsic curvatures $\k_{\pp}$. 

For the case of codimension 1 the junction conditions read,
\be
\label{jump}
-\sum_{k=0}^{\left [(D-1)/2\right ]} 8\alpha_k k! \,
 \left[ \sigma^{{\pp}}_{(N,k)\mu}\right] 
= 2 S_\mu{}^\nu (P)
\theta^{\star {\pp}}_\nu \, .
\ee
and the Dirac charge will be provided by the jump of $\sigma_{(N,k)\mu}^{\pp}$. Here $1\leq k \leq [(D-1)/2]$ and the higher the dimension the more the Lovelock charges that contribute.
For example in $p=3$ we get in turn for the $k=1$ Einstein and $k=2$ Gauss-Bonnet terms,
\ba
\label{boss}
\sigma^{\pp}_{(1,1)\mu} &=& 2 \, \k^{\nu}_{{\pp}N } \wedge \theta^{\star
{\pp}}_{\mu \nu} = - 2 \, (K^{\nu}_{\mu } - \delta^{\nu}{}_\mu K ) \theta^{\star {\pp}}_{\nu} \nonumber \\
\sigma^{{\pp}}_{(1,2)\mu} &=& 4 \, \k^{\nu}_{{\pp}N} \wedge \left
( \Omega^{\rho \lambda}_{{\pp}} - \frac{1}{3} \k^{\rho}_{{\pp}N} \wedge \k^{\lambda}_{{\pp}N} \right ) \wedge \theta^{\star {\pp}}_{\mu \nu \rho
\lambda} \nonumber\\
&=& -4 (3 J_\mu^\nu-J \delta_\mu^\nu -2 P^{\nu}_{\;\;\lambda\rho\mu}
K^{\lambda \rho}) \theta^{\star {\pp}}_{\nu} \, ,
\ea
where we have set
$$
J_{\mu\nu} = \frac{1}{3}(2K K_{\mu\lambda}K^\lambda{}_\nu + K_{\lambda\rho}K^{\lambda\rho} K_{\mu\nu} 
{}- 2K_{\mu\lambda}K^{\lambda\rho}K_{\rho\nu} - K^2 K_{\mu\nu}),
$$ 
$$
P_{\mu\nu\lambda\rho} = R_{\mu\nu\lambda\rho} + 2 R_{\nu[\lambda} g_{\rho]\mu} - 2 R_{\mu[\lambda}
g_{\rho]\nu} + R g_{\mu[\lambda}g_{\rho]\nu}
$$
and we have dropped the label $N$ for the extrinsic curvature components.
Replacing into the matching conditions (\ref{jump}) and taking the
corresponding left and right limits  one gets the
well-known junction conditions for Einstein \cite{chaisr} and Einstein-Gauss-Bonnet gravity
\cite{chadav}. The equations of motion can also be
derived from the boundary action, after variation with respect to the induced frame
$\theta^{\nu}$ on the left and right side of the brane. 
In differential form language the action is 
obtained trivially from (\ref{boss}) by removing the free index,
\ba
S_{\Sigma}=2\alpha_1 \int_{\Sigma} \k^{\nu}_{{\pp} N} \wedge \theta^{\star{\pp}}_{\nu}
+4 \alpha_2 \int_{\Sigma} \k^{\nu}_{{\pp}N} \wedge \left
( \Omega^{\rho \lambda}_{{\pp}} - \frac{1}{3} \k^{\rho}_{{\pp}N} \wedge
\k^{\lambda}_{{\pp}N} \right ) \wedge \theta^{\star {\pp}}_{\nu \rho \lambda}
\ea
and agrees with the Gibbons-Hawking \cite{chagib} and Myers \cite{chamye2} boundary
terms. 
It is obvious now that for $D=7$ a $5$-brane will have an extra term contributing to the junction conditions 
which will be of 5th order in the extrinsic curvatures etc. This can be understood intuitively since higher order Lovelock densities involve higher powers of curvature. For the case of higher odd codimension see \cite{chaCZ}.

Let us now concentrate on even codimension $N=2n$.
We will assume that the parallel sections are regular. Then when integrated out over $\Sigma_\perp$, the normal sections quite naturally yield 
the charge or topological defect of $(\chi-\beta) Area(S_{2n-1})$.
This is similar to the defect angle for the infinitesimal cosmic string
in 4 dimensions (\ref{chastring}) when we set $\chi=1$, where $\chi$ is the Euler characteristic of the normal section{\footnote{I thank Robin Zegers for pointing this out}}. In higher codimension one has to allow for non-trivial Euler character in $\Sigma_{\perp}$ in order to obtain  removable type  singularities \cite{chacod4}. Due to their topological character, we call these 
 topological matching conditions \cite{chaCZ}.
The matching conditions read,
\be
\label{chatopo}
(\chi-\beta) \mbox{Area}({S}_{2n-1}) \sum_{\tilde{k}=0}^{\left [{p\over 2}\right ]} 
\tilde{\alpha}_{{k}} \sigma^{{\pp}}_{(n,\tilde{k})\mu}(P) = -2 S_\mu^\nu (P)
\theta^{\star {\pp}}_\nu \, ,
\ee
where we have set $\tilde{\alpha}_{{k}}=2^{2n-1}k!(n-1)!\alpha_k/(k-n)!$ and the
smooth parallel forms $\sigma^{{\pp}}_{(n,k)\mu}$ are given by
\be
\label{chalemon}
\sigma^{{\pp}}_{(n,k)\mu}= \sum_{j=0}^{{\mbox{\footnotesize{min}}\left (n-1,
\tilde{k} \right)}}  \left( ^{\,\tilde{k}}_{j} \right )
\left (\bigwedge_{l=1}^{\tilde{k}-j} \R^{\lambda_{l} \nu_{l}}_{{\pp}} \right ) \wedge
\left ( \bigwedge_{l=1}^{2j} \k^{\rho_l}_{{\pp} N} \right ) \wedge
\theta^{\star {\pp}}_{\mu \lambda_{1} \nu_{1} \cdots \lambda_{\tilde{k} - j} \nu_{\tilde{k} - j}
  \rho_{1} \cdots \rho_{2j}} \, .
\ee
Finally, $\tilde{k}=k-n$, which ranges between $0\leq \tilde{k}\leq \left[{p\over 2}\right ]$, will turn out to be the induced Lovelock
rank of (\ref{chalemon}). The parallel forms (\ref{chalemon}) will dictate the
dynamics of the brane. Their expressions involve powers of the
projected Riemann tensor $\R_{\pp}$ on the brane and even powers of
the radial extrinsic curvature $\k_N$  of $\Sigma$. To see this, consider for
each $\tilde{k}$ the first term in the sum, $j=0$, which reads,
\be
\label{chamarion}
\sigma_{(\tilde{k},0)\mu}^{\pp}=\left (\bigwedge_{l=1}^{\tilde{k}} \R^{\lambda_{l}
    \nu_{l}}_{{\pp}} \right )  \wedge \theta^{\star {\pp}}_{\mu \lambda_1
    \nu_{1}...\lambda_{\tilde{k}}\nu_{\tilde{k}}} \, .
\ee
Note the similarity to the bulk Lovelock densities, (\ref{klov}). 
Clearly, using (\ref{chagform}), we see the appearance of induced Lovelock densities involving $\Omega_{\pp}$ accompanied by even powers of the extrinsic curvature $\k_N$. The former describe the induced quantites of the brane and the latter how $\Sigma$ is embedded in the bulk. In particular, for $\tilde{k}=0$ we will have a pure tension term, for $\tilde{k}=1$ an Einstein term and for $\tilde{k}=2$ a Gauss-Bonnet term with extrinsic curvature terms. Note that the highest rank on the brane originates from the highest rank Lovelock term in the bulk.  This agrees with the fact that in Einstein gravity, $k=1$, there are no matching conditions beyond $n>2$.

Let us compare our result with the work of \cite{chabos}. Thus, let us consider
the case of a 3-brane embedded in 6 dimensional spacetime, for which there are
only two terms in (\ref{chalemon}),
\ba
\label{chaeinscodim2}
\sigma_{(1,0) \mu}^{\pp} &=&  \theta^{\star {\pp}}_\mu, \qquad \sigma_{(1,1)
  \mu}^{\pp} = \R_{{\pp}}^{\nu \rho} \wedge \theta^{\star
{\pp}}_{\mu \nu \rho} \, . 
\ea
Thus, using (\ref{chaident}), (\ref{chacurv}) and (\ref{chagform}), we obtain from
  (\ref{chatopo}), 
\be
2 \pi \, (1-\beta)  \left \{ -\alpha_1 h_{\mu\nu}
 + 4 \alpha_2  G_{\mu\nu}^{(ind)}  - 4 \, \alpha_2
\,  W_{\mu\nu} \right \} 
=  S_{\mu\nu}  \, ,
\ee
where \cite{chabos}
\be
\label{w}
W_{\mu}^{\lambda}=K_{N} K^{\lambda}_{\mu N}  - K^\nu_{\mu N}
K^{\lambda}_{\nu N}  - \frac{1}{2}
\delta^{\lambda}{}_{\mu} \left( K_{N}^2 - K^\nu_{\rho N} K^{\rho}_{\nu N}
\right ) \, .
\ee
Note that the bulk Einstein term, $k=1$, 
only allows for an effective cosmological constant on the 
brane, while the
Gauss-Bonnet term, $k=2$, induces the Einstein tensor  
for the brane's equation of motion. This equation is  similar to the one
found in \cite{chabos}, the difference being that here,
the extrinsic geometry is supposed perfectly
regular. Indeed mathematically  there is no reason to suppose that the
extrinsic curvature has a jump, the topological defect carrying the necessary
charge in complete analogy to the case of the cosmic string. 
Aditionally, these matching conditions provide the maximal regularity
for the bulk metric. Furthermore, since the induced and extrinsic geometry are
smooth, the equations of motion can be seen to actually originate from a simple
action taken over $\Sigma$ as long as we suppose that $\beta=constant$. In other words, the degrees of freedom
associated to the normal section are completely integrated out giving an exact
action for the brane's motion. In differential form language, it is straightforward to read off the langrangian density
in question. Literally, taking out the free index for the charge in
(\ref{chamarion}) and using Gauss equation (\ref{chagform}), we get
\ba
\label{melina}
S_{\Sigma}^{(p=3,n=1)} &=& 2 \pi (1-\beta) \, \int_{\Sigma} \left( \tilde{\alpha_1} \theta^{\star \pp}+ \tilde{\alpha_2} (\R^{\nu
\rho}_{{\pp}} \wedge \theta^{\star {\pp}}_{ \nu\rho} )\right)+\int_{\Sigma}
\L_{matter} \nonumber\\ &=& 2 \pi (1-\beta) \int_{\Sigma} \sqrt{-h} \, \left (\tilde{\alpha_1}+ \tilde{\alpha_2}
(R^{ind}-K^2+ K_{\mu\nu}^2) \right ) +\int_{\Sigma}
\L_{matter} \, ,
\ea
We see appearing the cosmological constant, the Einstein-Hilbert term  with an extrinsic
curvature term quite similar to the finite width corrections one obtains for
cosmic strings in flat spacetime \cite{bonjour}!
In other words, the only bulk quantity entering in the equations of motion is
the topological defect fixing the overall mass scale 
and the extrinsic curvature of the surface giving a matter like component in
the action. Therefore the Gauss-Bonnet  term quite naturally gives an induced gravity term 
on the brane for codimension 2. This does not mean that there is a localised 0-mode graviton on the brane location but much like in DGP one can expect a quasi-localised 0-mode or in other words ordinary 4 dimensional gravity up to some cross-over scale as in the DGP model \cite{chadgp}. It would seem that the topological quantity $\beta$ would be giving us in this case a crossover scale for 4 dimensional gravity to 6 dimensional gravity. Clearly perturbation theory of well-defined warped backgrounds is needed in order to answer this question.

Equation (\ref{chatopo}) has another surprising property. Indeed,  an important
simplification takes place if we suppose that 
\be
\label{isa}
\left[ p/2\right] +1 \leq n \, ,
\ee 
{\it ie} that the codimension of the brane is larger than its 
intrinsic dimension. In that case, using Gauss's equation, we have
\be
\label{simple}
\sigma^{{\pp}}_{(n,k)\mu}= \left (\bigwedge_{l=0}^{\tilde{k}} 
\Omega^{\lambda_{l} \nu_{l}}_{{\pp}} \right ) \wedge 
\theta^{\star {\pp}}_{\mu \lambda_1 \nu_{1} \cdots \lambda_{\tilde{k}} \nu_{\tilde{k}}} 
\ee
and the matching conditions (\ref{chatopo}) are simply the induced Lovelock
equations on the brane with no extrinsic curvature terms! Therefore, the action
for a distributional 3-brane embedded in $D=8,10,... 2d$ dimensions is exactly
the Einstein-Hilbert plus cosmological constant action
$$
S_{\Sigma}^{(p=3,n>1)}= (1-\beta) \mbox{Area}(S_{2n-1}) \int_{\Sigma} \sqrt{-h} \left
  ( \tilde{\alpha}_n+ \tilde{\alpha}_{n+1}
R^{ind} \right )+\int_{\Sigma} \L_{matter} \, ,
$$
with Planck scale set by
$$
M_{Pl}^2=(1-\beta)\, \mbox{Area}(S_{2n-1})\, \tilde{\alpha}_{n+1} \, .
$$
For a 4 or 5-brane, we will have in addition the Gauss-Bonnet term etc. In other
words, if the codimension verifies (\ref{isa}), all the extrinsic curvature
corrections drop out and there is a complete Lovelock reduction from the bulk
to the induced Lovelock terms on the brane.   All the degrees of freedom originating from the bulk at zero
thickness level are exactly integrated out giving the most general classical equations of motion for the brane. 
As a consequence we have in particular energy-conservation on
the brane (see also \cite{chakof1}).

\section{Applications to braneworlds}

Braneworlds offer interesting and direct applications to Lovelock theory. This is due to the fact that Lovelock theory is the most general metric theory that can enjoy well defined junction conditions  There are two reasons for this. Firstly, the field equations beeing of second order Dirac matter distributions still allow a continuous metric across the junction surface and therefore contraints are imposed on geometric quantities such as the extrinsic curvature and the induced curvature of the braneworld. They will lead inevitably to a self gravitating equation of motion for the surface in codimension 1 and in  interesting contstraints for codimension 2. Furthermore, in higher codimension \cite{chaemparan} Einstein theory does not pocces the necessary richness in order to admit distributional sources, rather the finite thickness of the defect plays a neccesary role or one has to admit the presence of non-removable curvature singularities at the brane location. Lovelock theory seemingly admits higher codimension distributions \cite{chazeg} and simple examples of codimension 4 braneworlds are emerging \cite{chacod4}. Here we will concentate in turn to codimension 1 and codimension 2 braneworlds.  

\subsection{Codimension 1 braneworlds and their effective 4d gravity}

In this section{\footnote{I thank Stephen Davis for unpublished collaboration throughout this section which in part can be found in his papers \cite{chadav2}}} we will consider the application of Lovelock theory to a 4 dimensional braneworld embedded in a 5 dimensional curved backround spacetime. We will take  a flat single brane in adS ie a Randall-Sundrum type braneworld and its perturbations and we will then give the relevant cosmological evolution equations. We will also include the induced gravity term so as to consider the most general 5 dimensional configuration with a brane boundary. 
For simplicity we will also consider $Z_2$ symmetry although the assymetric cases present a number of interesting features \cite{chapad2}. 
Consider therefore the action,
\ba
\label{chaaction2}
&&S=\frac{M^3}{2} \int_\M d^5x \sqrt{-g} \,
\left(\zeta R +\hat{\alpha} \hat{G} - 2\Lambda\right)
\nonumber \\  &&\hspace*{0.3 in} 
{}-M^3\int_\Sigma d^4x \sqrt{-h} \, \left[\zeta K 
+ 2\hat{\alpha} (J-2 \widehat G^{ab}K_{ab}) \right]
\nonumber \\  &&\hspace*{0.3 in}
{}+ \frac{M^3}{2}\int_\Sigma d^4x \sqrt{-h} \, \left(\beta \widehat R - 2T\right)
-\int_\M d^5x \sqrt{-g} \, \L_{mat}-\int d^4x \sqrt{-h} \, \widehat \L_{mat}
\ea
where the first line represents the bulk Lovelock theory, the second the Gibbons-Hawking and Myers boundary terms and the third the induced gravity and matter contributions. Notice that we have included a dimensionless number $\zeta$ in front of the Einstein-Hilbert term since we will take its zero limit in order to see the resulting 4 dimensional brane gravity of a pure Gauss-Bonnet bulk term. Given the normal vector $n_a$, the 5 dimensional projector 
\be
\label{chafund}
h_{ab} = g_{ab} - n_a n_b
\ee
is the surface induced metric and
\be
\label{chafund2}
K_{ab} = h^c{}_a \nabla_{\! c} n_b
\ee
is the extrinsic curvature (strictly speaking the first and second fundamental forms of 
$\Sigma$). The caret denotes tensors constructed out of $h_{ab}$.  $[X]$ as before denotes the jump in
$X$ across the brane.

The field equations in the bulk and on the brane are,
\be
\label{chafield}
\zeta G_{ab}+ \hat{\alpha} H_{ab} + \Lambda g_{ab} = M^{-3} T_{ab}
\ee
where $H_{ab}$ is is the second order Lovelock tensor (\ref{chalov23}). The bulk and brane energy-momentum
tensors are respectively $T_{ab} = 2\delta \L_{mat}/\delta h^{ab} - h_{ab} \L_{mat}$ 
and $S_{ab} = 2\delta \widehat \L_{mat}/\delta h^{ab} - h_{ab} \widehat\L_{mat}$. 
According to (\ref{jump}) \cite{chadav} we have 
\be
\zeta\left[K_{ab} - K h_{ab}\right] 
+ \alpha\left[ 3 J_{ab} - J h_{ab} +  2\widehat P_{acdb} K^{cd} \right]
- \beta \widehat G_{ab} = -M^{-3} S_{ab} + T h_{ab}
\ee
For an adS bulk and a flat brane situated at $z=0$ 
we use a Poincar\'e patch and the solution reads,
\be
\label{chars2}
ds^2 = a^2(z) \eta_{\mu\nu} dx^\mu dx^\nu + dz^2
\ee 
where the warp factor is $a(z)=e^{-k|z|}$ and we have,
\be
\Lambda = -6 k^2 (\zeta-2\alpha k^2) \ \ \ , \ \ \ 
T = 2k(3\zeta-2\alpha k^2)
\ee
for the bare cosmological constant and the brane tension respectively. The effective curvature scale $k$ is given by
$2\alpha k^2=\zeta\mp\sqrt{\zeta^2+ 2 \alpha \Lambda/3}$ from solving the field equations in the bulk and on the boundary. The upper sign corresponds to  the Einstein branch. 

To obtain the effective four-dimensional gravity induced on the brane, we consider a general linear perturbation theory parametrised by $\gamma_{ab}(x,z)$, around the background solution (\ref{chars2}), 
\be
ds^2 = a^2(z)(\eta_{\mu \nu} + \gamma_{\mu \nu}) dx^\mu dx^\nu 
+ 2 \gamma_{\mu z} dx^\mu dz + (1+\gamma_{zz}) dz^2
\label{chapertmet}
\ee
We will deal with gauge in a moment.
We consider in particular the perturbed four dimensional Ricci tensor corresponding to
the linear perturbation $\gamma_{\mu\nu}$,
\be
\mathcal{R}_{\mu\nu} =  \frac{1}{2}\left(
2\partial^\alpha \partial_{(\nu} \gamma_{\mu)\alpha}
- \Box_4 \gamma_{\mu\nu} - \partial_\mu \partial_\nu \gamma\right)
\ee
and  we define $\mathcal{R} = \eta^{\mu\nu}\mathcal{R}_{\mu\nu}$ and 
$\mathcal{G}_{\mu\nu} = \mathcal{R}_{\mu\nu} - (1/2) \eta_{\mu\nu}\mathcal{R}$. These  are the typical geometrical quantities encountered in perturbation theory of induced gravity terms such as DGP braneworlds \cite{chadgp}.

Given our mirror symmetry bulk equations (for $z>0$) are then
\be
(\zeta-2\alpha k^2)\left\{(6k \partial^\mu \gamma_{z \mu} - \mathcal{R}) e^{2k z}
-3 k(4k \gamma_{zz} + \dz \gamma)\right\} = \frac{2}{M^3} T_{zz}
\label{gauss}
\ee
\be
(\zeta-2\alpha k^2)\left\{ 
(\partial_\mu \partial^\nu \gamma_{z \nu} - \Box_4 \gamma_{z \mu})e^{2k z}
-3 k\partial_\mu \gamma_{zz} - \dz(\partial^\nu \gamma_{\mu\nu} - \partial_\mu \gamma)
\right\} = \frac{2}{M^3} T_{z\mu}
\label{codacci}
\ee
and 
\ba
(\zeta-2\alpha k^2)\left\{ \left(2\mathcal{G}_{\mu\nu} 
+ (\eta_{\mu\nu} \Box_4 - \partial_\mu \partial_\nu)\gamma_{zz}
-2(\dz -2k)(\eta_{\mu\nu} \partial^\alpha v_\alpha - \partial_{(\mu} v_{\nu)})
\right)e^{2kz} 
\right. \hspace{.5in} \nonumber \\ \left. {}
- (\dz^2-4k \dz)(\gamma_{\mu\nu} - \eta_{\mu\nu} \gamma)
+3k\eta_{\mu\nu}(\dz -4k)\gamma_{zz}\right\} = \frac{2}{M^3} e^{2kz}T_{\mu\nu}
\ea
The higher order Gauss-Bonnet contribution is recognised easily by the appearence of the coupling constant $\alpha$ while the usual Einstein terms are identified by $\zeta$.
Note first of all how the metric perturbation in the bulk are quasi-identical to those in Einstein theory apart from the overall coupling of $\alpha$ with the warp factor $k$ multiplying the differential operator. If there were no warp factor, for example in a Minkowski bulk, the higher order Gauss-Bonnet term would have had {\it{ no contribution in linear perturbation theory}}. A general remark we can make is that since the coupling constant $\alpha$ is dimensionful $\alpha \sim [length]^2$ needs inevitably a bulk curvature scale to couple to. This can be related for example to a bare cosmological constant in the bulk or an effective cosmological constant on the brane \cite{chafax2}. {\footnote{When spacetime is not conformally invariant as for example for a black hole then the perturbation operator picks up extra terms related to the background Weyl curvature that appears in the field equations for Lovelock theory for $k \geq 2$!}} Also note that given our couplings to matter in action (\ref{chaaction2}) we need to have $\zeta-2\alpha k^2\geq 0$.  We otherwise quite clearly have a bulk ghost. If we switch off Einstein gravity in the bulk $\zeta=0$ and choose $\alpha<0$ we have Einstein like perturbation theory.  When $\zeta=2\alpha k^2$ the linear perturbation operator switches off and seemingly we approach a strongly coupled limit. This is the case of Chern-Simons gravity which is very interesting by itself and we invite the interested reader to consult \cite{chazan}.
In turn the junction conditions at the brane $z=0$ are 
\be
\left[(\zeta-2\alpha k^2)\left\{3k\eta_{\mu\nu} \gamma_{zz} 
- 2\eta_{\mu\nu} \partial^\alpha v_\alpha + 2\partial_{(\mu} v_{\nu)} 
- \dz (\gamma_{\mu\nu} - \eta_{\mu\nu} \gamma)\right\}\right]
+ 2(\beta+4\alpha k) \mathcal{G}_{\mu\nu} = \frac{2}{M^3} S_{\mu\nu}
\ee
Here note that the higher order Lovelock term give two contributions. Firstly a Neumann type boundary term in the sense that it involves the first derivative of the perturbation metric and is exactly the same as the usual Einstein perturbation. Secondly, an induced gravity type term, which is accompanied by the induced gravity coupling $\beta$ and which yields an induced Einstein term on the brane.  
Note that indices in the above equations are raised/lowered with
$\eta^{\mu\nu}$, and that 
$T^a{}_a = T_{zz} + e^{-2kz}\eta^{\mu\nu} T_{\mu\nu}$. 

Before solving the above equation let us first deal with gauge freedom. Any infinitesimal bulk transformation,
$x^a\rightarrow x^a +\zeta^a$ gives the transformation law $\gamma_{ab}\rightarrow \gamma_{ab}+\L_\zeta g_{ab}$ which leaves the bulk equations invariant. Here take the infitesimal shift,
\ba
z\rightarrow z+\epsilon(x,z)\nonumber\\
x^\mu \rightarrow x^\mu+\xi^\mu(x,z)+\partial^\mu \xi(x,z)
\ea
(where $\partial_\mu \xi^\mu=0$) and obtain the linear isometries,
\ba
h_{\mu y}\rightarrow h_{\mu y}-a(z)( \xi'_\mu+\partial_\mu \xi')-\partial_\mu \epsilon\\
h_{y y}\rightarrow h_{y y} -2\epsilon'\\
h_{\mu \nu} \rightarrow h_{\mu \nu}-\partial_\mu \xi_\nu-\partial_\nu \xi_\mu-2\partial_\mu \partial_\nu \xi -2 a'(z)\epsilon \eta_{\mu\nu}
\ea
which leave the bulk equations invariant. A bulk coordinate transformation however can also displace the brane by,
$F \rightarrow F + \epsilon(x,z=0)$ where we call $F$ the brane bending mode. 
This is avoided as long as we choose $\epsilon(x,z=0)=0$ on the brane.

It is easy now to solve  the equations (\ref{gauss}) and (\ref{codacci}) when $T_{ab}=0$
and $k \neq 0$ by going to a fixed wall gauge where there is no brane bending $F$ and the wall is maintained at $z=0$.
In this gauge the brane boundary conditions are also invariant under the bulk transformations \cite{hol}. We get,
\be
\gamma_{zz}= -\frac{1}{4k} \, \dz \gamma 
\label{gauge01}
\ee
\be
\gamma_{z \mu} = -\frac{\sgn(z)}{8k} \, \partial_\mu \gamma + B_\mu
\label{gauge02}
\ee
\be
\partial^\mu \bar \gamma_{\mu \nu} = 0
\label{gauge03}
\ee
where $\bar \gamma_{\mu \nu} = \gamma_{\mu \nu} - (1/4) \gamma \eta_{\mu \nu}$ is the trace-free part of $\gamma_{\mu\nu}$
and the trace reads $\gamma = \eta^{\mu \nu}\gamma_{\mu \nu}$.
We also have $\partial^\mu B_\mu= 0$ and $\Box_4 B_\mu=0$ but we can choose gauge for which  $B_\mu=0$ as we will do from now. 

The remaining bulk field equation is given by
\be
(\zeta - 2\alpha k^2)\left(\dz^2 -4k \dz + e^{2k z}\Box_4\right) 
\bar \gamma_{\mu\nu}=0
\ee
for $z>0$.
The boundary conditions at $z=0$ give
\be
2(\zeta-2\alpha k^2)\dz \bar \gamma_{\mu\nu}
+ (\beta + 4\alpha k) \Box_4 \bar \gamma_{\mu\nu}
=-\frac{2}{M^3}\left\{S_{\mu\nu} - \frac{1}{3} 
\left(\eta_{\mu\nu} - \frac{\partial_\mu \partial_\nu}{\Box_4}\right) 
S \right\}
\label{pertbc}
\ee
and
\be
(\zeta + \beta k + 2\alpha k^2) \Box_4 \gamma = \frac{4k}{3M^3}S
\ee
It becomes clear here that, as we mentioned above, the equivalent of `brane-bending' effects are included in the tracefull part of the metric $\gamma$ which is a genuine scalar mode since it couples to matter.  Following Davis \cite{chadav} (see also \cite{chafax2}, \cite{chamei})
we obtain the bulk solution (for $p^2>0$) which vanishes as $z\to \infty$,
\be
\bar \gamma_{\mu\nu}(p,z) \sim e^{2k|z|} 
K_2 \left(\frac{|p| e^{k |z|}}{k}\right)
\ee
for $k>0$, where $K_2$ are special Bessel functions of the second kind.
Therefore the boundary conditions imply
\be
\bar \gamma_{\mu\nu} = 
\frac{e^{2kz} K_2 (p e^{k|z|}/k)}{K_2(p/k)}
\frac{2k}{p^2F(p/k) M^3}
\left\{S_{\mu\nu} - \frac{1}{3}\left(\eta_{\mu\nu} - \frac{p_\mu
p_\nu}{p^2}\right) S\right\}
\ee
where
\be
F(p/k) = 2(\zeta-2\alpha k^2)\frac{ K_1(p/k)}{p/k K_2(p/k)} 
+ (k \beta+4\alpha k^2) 
\ee
and
\be
\gamma = -\frac{k}{p^2} \frac{4}{3(\zeta + \beta k + 2\alpha k^2)M^3} S
\ee
Any zeros of $F$ with respect to momentum will correspond to tachyon modes on the brane (see \cite{chafax2} for the 2 brane case) and this is essentially due to the fact that we have mixed boundary conditions.
The condition for a perturbatively stable theory is to have no
ghosts (brane or bulk) and no tachyons, therefore, $\zeta-2\alpha k^2>0$ and 
$\beta k+4\alpha k^2 \geq 0$. 

We now keep $k$ fixed.
For large distances and very small momenta $p/k$
\be
\label{small} 
F(p/k) = (\zeta+k\beta+2\alpha k^2) +
\frac{(\zeta-2\alpha k^2)}{2k^2} p^2
\left(\ln\frac{p}{2k}+\gamma_E\right) + O(p^4/k^4)
\ee
where $\gamma_E \approx 0.577$ is Euler's constant.
For small distances and very large momenta $p/k$ we have
\be
F(p/k) = (\beta k + 4\alpha k^2)
 + (\zeta-2\alpha k^2)\left(\frac{2k}{p}
- \frac{3k^2}{p^2}+O((p/k)^{-3})\right) 
\ee

Hence for large distances (small $p$)
\be
\label{1}
\mathcal{G}_{\mu\nu} 
= \frac{k}{(\zeta+k\beta+2\alpha k^2)M^3} S_{\mu\nu} + O(p^2)
\ee
we obtain Einstein gravity with $\mpl^2 = M^3 (\zeta+k\beta+2\alpha k^2)/k$. This is true even if $\zeta=0$

For the short distance behavior and large $p$,  we define an effective scalar mode
\be
\phi = -\frac{\zeta- 2\alpha k^2}{2k(\beta+4\alpha k)} \, \gamma
\ee
then we obtain
\be
\mathcal{G}_{\mu\nu} -2(\eta_{\mu\nu}\Box_4 - \partial_\mu \partial_\nu)\phi
=\frac{1}{(\beta+4\alpha k)M^3} S_{\mu\nu} + O(p^{-1})
\ee
and
\be
\Box_4 \phi = 
-\frac{2(\zeta- 2\alpha k^2)}{3(\zeta + \beta k + 2\alpha k^2)}
\frac{1}{(\beta+4\alpha k) M^3}S
\ee
which is linearised Brans-Dicke gravity with 
$\mpl^2 = M^3(\beta+8\alpha k)$ and 
\be
(2\omega+3) = \frac{3(\zeta + \beta k + 2\alpha k^2)}{4(\zeta- 2\alpha k^2)}
\ee
$\omega$ a Brans-Dicke coupling. This is clearly an effect absent in conventinal GR and is present due to the Gauss-Bonnet term and the induced gravity term on the brane parametrised by $\alpha$ and $\beta$ respectively.
Solar system constraints dictate that $\omega>4 \;10^4$ in order to agree with time delay experiments as that of the Cassini spacecraft, \cite{chacasini}.  Therefore $\omega$ should be pretty big and hence we need to have $\zeta- 2\alpha k^2 \approx 0$, close to the Chern-Simons case.

\subsection{Codimension 1, brane cosmology}

Having looked at the perturbation theory of ra codimension 1 braneworld let us now study the exact solution for cosmology (see also Gregory's lecture notes \cite{charuth0}). Consider   a
4-dimensional cosmological 3-brane of induced geometry,
\be
\label{chainduced}
ds^2_{ind}=-d\tau^2+R^2(\tau)\left(\frac{d\chi^2}{1-\kappa \chi^2} +\chi^2 d\Omega_{II}\right)
\ee
 We suppose, following the symmetries of our metric,
that matter on
the brane is modelled by a perfect fluid of energy density $\E$ and
pressure $\P$. The brane is fixed at $z=0$,
and the energy-momentum tensor associated with the brane takes the form,
$$
S_{\mu}^{\nu (b)} =\mathrm{diag}(-\E(\tau),\P(\tau),\P(\tau),\P(\tau)).
$$
We assume $Z_2$ symmetry across the location of the brane at $z=0$,
and set $M^{3}=1$ for the time being. The bulk symmetries are 4 dimensional cosmological symmetries imposed from (\ref{chainduced}). In other words we have three dimensional  isotropy and homogeneity which leads us to the bulk metric (\ref{chametric1}). Using (\ref{chametric1}) and solving the field equations in the bulk we found two {\it{bulk}} gauge degrees of freedom, $U$ and $V$. These, leave the brane junction conditions invariant, or equivelantly, the distributional part of the field equations (\ref{chalov23}) invariant, if and only if $U=V$. The one remaining bulk-brane physical degree of freedom can be traced after coordinate transformation to the expansion factor or brane trajectory $R=R(\tau)$ \cite{chapeter}, \cite{chafax1} evolving in the black hole bulk (\ref{BH}). At the end of the day  setting,
$$
\A=H^2+{\kappa\over R^2}
$$
where $H=\frac{\dot{R}}{R}$ the generalised Friedmann equation reads,
\ba
\label{chafried1} 
\E-3\beta \A = \sqrt{{2\A\alpha + \zeta - U \over \alpha}}2(4\A \alpha +2\zeta
+ U)
\ea
where we have defined
$$
U = \pm \sqrt{\zeta^2+{2 \alpha \Lambda\over 3} + \frac{4\alpha\mu}{R^4}}
$$
and we set 
\be
U_0 = U(\mu=0) = \zeta - 2\alpha k^2
\ee
Thus to avoid bulk ghosts we must have $U_0>0$, and so the lower branch is
ruled out.
The standard
conservation equation on the brane remains valid and reads,
\be
\label{chaconserv}
\frac{d\E}{d\tau}+ 3 H (\P+\E) =0
\ee

Squaring (\ref{chafried1}) gives a third order polynomial for $\A$,
\be
\label{chafried2} 
{\E^2\over 2\alpha}=32\alpha \A^3
+{3\over 2\alpha}\A^2(32\alpha\zeta-3\beta^2)
-{3\over \alpha}(2U^2-\E\beta-8\zeta^2)\A
-{1\over 2\alpha^2}(U-\zeta)(U + 2\zeta)^2 \ .
\ee

Setting $\E=T+\rho$ and $\mu=0$ we can read off the critical
tension from (\ref{chafried2}).
Critical means that the effective cosmological constant on the
brane is zero,
\be
\label{tension}
T_c^2=2{\zeta- U_0\over \alpha}(2\zeta + U_0)^2 = 4k^2(3\zeta-2\alpha k^2)^2
\ee
This condition has to be imposed if we want to analyse genuine geometric effect on the modified Friedmann equation. 
In turn this gives to linear order in $\rho$
the reduced four-dimensional Planck mass
\be
\label{mplanck}
\frac{1}{\mpl^2} = \frac{1}{M^{3}} {\sqrt{\zeta- U_0}\over
 \sqrt{2\alpha}(2\zeta - U_0) +\beta \sqrt{\zeta- U_0}}
\ee
where we restore the fundamental mass scale $M$
of the 5-dimensional theory. This agrees with 
(\ref{1}) obtained in the previous section.

Before proceeding to analysing specific cases it is useful to recast the
Friedmann equation inputing $T_c$ and the warp factor $k$. In order to
avoid ghosts and tachyons we have $\beta> -8\alpha k/3$ 
so let us set $\beta=\beta_0-8\alpha k/3$.  Then
(\ref{chafried1}) for $\kappa=0$ reads,
\be
\E = 3\beta_0 H^2- 8\alpha k(1-\Q)H^2+(1-\M) {8\alpha k^3\over
3}\Q+{2\Q\over 3}T_c (1+{\M\over 2})
\ee
where 
$$
\Q=\sqrt{{H^2\over k^2} + {1+2\M \over 3} +{T_c\over 12\alpha k^3}(1-\M)}
$$
and 
$$
\M=\sqrt{1+{4 \alpha \mu\over R^4 ({T_c\over 6k}-{4\alpha k^2\over 3})^2}}
$$
Now we see that in accord with perturbation theory the Gauss-Bonnet term yields an ordinary Friedmann 
term just as in induced gravity. This is parametrised by $\beta_0$. We emphasize that this term is due to the higher order Lovelock correction and not due to the higher order Einstein-Hilbert term. Therefore a naive expectation that an Einstein-Hilbert term of higher dimension gives 4 dimensional gravity whereas a Gauss-Bonnet term differing phenomenology is not true. Quite the contrary, higher order Lovelock terms even in codimension 1 give naturally ordinary 4 d gravity at some scales. This can be also expected by the presence of the induced gravity term in the Myers boundary for Gauss-Bonnet theory {\footnote{I thank Nemanja Kaloper for pointing this out}}. 
Starting from the above we can obtain the second FRW
equation which will tell us about acceleration.

As an example let us now consider the case of zero tension. For a general analysis see also \cite{charoy}. Although this means that we fine tune the couplings, 
$2\alpha k^2 = 3\zeta$ in essence nothing special happens and the equations become easier to deal with. This means that in order not to have ghosts we take $\alpha<0$, upon which $\beta_0\sim
M_{pl}^2$ (for M=1). The Friedmann equation (for $\kappa=0$)  simplifies to,
\be
\label{nobrain}
\rho = 3\beta_0 H^2+ 8|\alpha| k(1-\Q)H^2-(1-\M) {8|\alpha| k^3\over 3}\Q
\ee
where now 
$\Q=\sqrt{{H^2\over k^2} + {1+2\M \over 3}}$ 
and $\M=\sqrt{1-{9\mu\over 4|\alpha| k^4 R^4}}$
with $\M=1$ for a pure adS background. For late time acceleration we
need the second FRW equation. To obtain it, 
we differentiate (\ref{nobrain}) and successive 
use of (\ref{chaconserv}) and (\ref{nobrain}) gives,
\ba
\label{accel}
&-&(\rho+3P)+{8|\alpha| k^3\over \Q}
\left( {H^2\over k^2} + {1-\M \over 3}\right)
\left(-{H^2\over k^2} + {2(1-\M^2) \over 3\M}\right)-{16(1-\M) \Q\over
3\M}|\alpha| k^3 = \nonumber \\
&=& 2{\ddot{R}\over R}\left[{\rho\over H^2}+{8|\alpha| k^3\over
3H^2}(1-\M) \Q - {4|\alpha| k\over \Q} 
\left({H^2\over k^2} + {1-\M \over 3}\right) \right]
\ea

Setting $\alpha=0$ gives ordinary Friedmann equations. This limit
effectively kills all the 5 dimensional part of the action for the special case we are treating. 
The second term
in (\ref{nobrain}) can be interpreted as an $\alpha$ correction (of
negative sign) and the third term a bulk black hole term (of positive
sign). The terms on the LHS of (\ref{accel}) tell us whether there is
possible acceleration or not at late time for ordinary matter ($w=0$) or
radiation $w=1/3$. This fact is true 
 modulo the sign of the parenthesis of the RHS
which may also change sign as, we will see, inversing the accelerating
equations of state but breaking the strong energy condition given (\ref{chaconserv}). 

It is now indicative to set $\mu=0$ and study late-time effects. Then
(\ref{nobrain}) simplifies to, 
\be
\label{indeed}
\rho = 3\beta_0 H^2 + 16|\alpha| H^2(k- \sqrt{k^2 + H^2})
\ee
and (\ref{accel}) to,
\be
\label{typical}
-(\rho+3P)-{8|\alpha| H^4 \over \sqrt{k^2+ H^2}}
= 2{\ddot{R}\over R}\left[{\rho\over H^2}- 
{4|\alpha| H^2\over \sqrt{H^2+k^2}} \right]
\ee  
The first term on the RHS of (\ref{indeed}) is the usual FRW term (with the
right Planck mass) 
but the second term is
of negative sign which means that we cannot take $H$ arbitrarily large  For late times cosmology approaches usual LFRW cosmology.
Lastly let us set the bulk Einstein-Hilbert term to zero $\zeta=0$, and take $\alpha<0$. Then $T_c=-4\alpha k^3$ and 
$$
\Q=\sqrt{{H^2\over k^2} + \M}
$$
with
$$
\M=\sqrt{1+{\mu\over \alpha k^4 R^4 }}
$$
Then the effective LFRW equations reduces to, 
\be
\E = 3\beta_0 H^2- 8\alpha k(1-\Q)H^2-4\alpha k^3 \M \Q
\ee
which gives at late times ordinary 4 dimensional cosmology despite the fact that there is no Einstein-Hilbert in the bulk action. This proves our claim in the introduction that the higher order terms of Lovelock theory enhance ordinary 4 dimensional gravity.

\subsection{Codimension 2 braneworlds}

Let us now look at a braneworld of codimension 2, in other words let us consider 6 dimensional bulk spacetime with 4 dimensional maximally symmetric subsections.  
The general bulk solutions, as we saw in section 2,  are solitons of  manifest axial symmetry with $\partial/\partial\theta$ as the angular Killing vector,
\be
\label{chasol}
ds^2=V(r) d\theta^2+\frac{dr^2}{V(r)}+r^2 d^2 K_4
\ee
and $d^2 K_4$ is the 4 dimensional line element of adS, flat or dS spacetime, $\kappa=-1,0,1$ respectively. For suitable parameters in (\ref{chapot1}) the radial coordinate varies inbetween $r_-\leq r \leq r_+$, where $r=r_{\pm}$ are the former horizon positions for (\ref{chabh}) and will be the possible brane locations, $r=r_\pm$ for (\ref{chasol}). 
In particular $r=r_+$ can be infinity itself in which case the effective volume element in the $(r,\theta)$ direction is infinite (for the analysis in the Einstein case see \cite{chamuko}). This is the codimension 2 version of warped compactification. We see therefore that  the six dimensional soliton (\ref{chasol}) posseses the correct spacetime symmetries to describe the general maximally symmetric 4 dimensional braneworld of constant curvature. The Wick rotated version of the staticity theorem tells us in particular that axial symmetry comes for free and need not be imposed when resolving the system of bulk equations. In order to introduce codimension 2 branes carrying some 2 dimensional Dirac charge or tension we need to reintroduce conical singularities at the relevant axis origins (\ref{chacyl}) which are at $r=r_h=r_-$ and $r=r_a=r_+$ by allowing for the presence of deficit angles $\beta_\pm$.  Indeed from (\ref{chareg}) we have the identification $\theta \sim \theta+\frac{2\pi-\beta_\pm}{\half V_{\pm}'}$ which means that we have the relation,
\be
\label{chaarni}
\frac{\beta_+}{\half V_+'}=\frac{\beta_-}{\half V_-'}
\ee
which relates the topological parameters $\beta_{\pm}$ with the geometrical quantities such as mass, charge of the soliton metric (\ref{chasol}). In particular note that we can always get rid of one of the conical singularities (and the resulting tensionful brane) and thus construct warped spacetimes with finite volume element and a single brane.  
Since the extrinsic curvatures for the branes are zero the brane junction conditions \cite{chabos} are given by,
\be
2\pi (1-\beta_\pm)\left(\delta_\mu^\nu +4\alpha G_{\mu \mbox{ind}}^\nu \right)=S_\mu^\nu
\ee
with $T_{\mu\nu}^{brane}=S_{\mu\nu} \delta^{(2)}(\rho)=S_{\mu\nu} \frac{\delta(\rho)}{2\pi \rho}$. We see the appearence of the induced Einstein tensor on the brane originating from the Gauss-Bonnet bulk term in the Lovelock action \cite{chazeg}. Note that the warp factor is given by the value of $r^2_{\pm}$. and in particular for $\kappa=1$ say $G_{\mu \nu}^{ind}=-3 H_{\pm}^2 \gamma_{\mu\nu}=-\frac{3}{r_{\pm}^2} \gamma_{\mu\nu}$ where $\gamma_{\mu\nu}$ is the de sitter induced metric with curvature set to 1. The induced Newton's constant on the brane is given by 
\be 
G^\pm_4=\frac{G_6}{ 8\pi \hat{\alpha}(1-\beta_\pm)} \label{chantNewt}
\ee 
A complete analysis of these solutions is given in  \cite{chaantonis} where self-tuning and self accelerating solutions are studied.

\section{The extended Kaluza-Klein reduction}

In the previous section we discussed warped compactifications of higher dimensional spacetimes. In this section we will discuss the case of Kaluza-Klein cosmpactification.
It is well known that the Kaluza-Klein (KK) reduction of Einstein theory gives us an Einstein-Maxwell-dilaton theory (EMD) for  specific KK couplings and some periodic boundary conditions. But what is the resulting theory  for the KK reduction of a higher dimensional Lovelock theory? Is the resulting reduced theory going to give second order field equations? Is the resulting theory unique as its higher dimensional counterpart?
Let us take here for simplicity the Kaluza-Klein reduction of an Einstein-Gauss-Bonnet theory from $D=5$ dimensions to $d=4$ dimensions. The full analysis in arbitrary $D$ and for Lovelock theory was carried out in \cite{chamul} (see also \cite{chamad} for applications to cosmology).
Starting with the relevant Lovelock theory in 5 dimensions, 
\be
\label{chaaction}
S=\frac{1}{16\pi G} \int_\M d^4x \, dy\, \sqrt{-g^{(5)}}\left(R^{(5)}-2\Lambda + \alpha \hat{G}^{(5)} \right)
\ee
we consider the following Anzatz for the 5 dimensional metric,
\ba
\label{chaKKmetric}
ds^2 &=& g_{ab}^{(5)} dx^a dx^b \nonumber\\
&=&(g_{\mu\nu}+e^{-4\phi}A_{\mu} A_{\nu})dx^\mu dx^\nu+2 A_\mu e^{-2\phi} dx^\mu dy+e^{-4\phi} dy^2
\ea
As usual we are making the basic assumption that there is a Killing vector $\partial_y$ in the 5th direction, in other words we do not consider here warped solutions we are rather interested in integrating out directly and obtaing the resulting  4 dimensional theory.  We impose periodic boundary conditions on $y$. We expect that the resulting 4 dimensional theory will be an extension of the Einstein Maxwell dilaton (EMD) theory in 4 dimensions with some exponential potential (in the presence of a 5 dimensional cosmological constant). Indeed integrating out the $y$ direction we obtain  \cite{chamul},
\ba
\label{chaKK}
S_{\mbox{eff}}&=&\int_{\M_4}d^4 x \, \sqrt{-g} e^{-2\phi} \left\{R-(\nabla \phi)^2-2\Lambda e^{\gamma \phi}-\quarter F^2 +\alpha \hat{G}^{(4)} + \right.\nonumber \\ 
&+& \left.\frac{3\alpha}{16} e^{-8\phi} \left[\left(F_{\mu\nu} F^{\mu\nu}  \right)^2-2 F_\mu^\nu F_\nu^\lambda F_\lambda^\kappa F_\kappa^\mu \right] \right\}-S_{int}
\ea
where note the presence of a non trivial interaction term which reads,
\be
\label{chainter}
S_{\mbox{int}}=-\half\int d^4x \sqrt{-g} e^{-6\phi}\left(F_{\mu\nu} F^{\kappa \lambda} R^{\mu\nu}_{\; \kappa\lambda} -4 F_{\mu\kappa} F^{\nu\kappa} R^\mu_{\;\nu} - F^2 R \right)
\ee
The interesting result that one can prove is that the field equations obtained from variation of the fields are still of second order as for the higher dimensional Lovelock metric theory. In fact the higher order EMD theory in question is the most general second order  theory that has up to second order partial derivatives. Any other numerical combination of the interaction terms for example would have given higher order derivatives. Hence we come to the interesting conclusion that higher dimensional Lovelock theories when dimensionally reduced via the KK formalism retain their  nice properties dictated by Lovelock's theorem. 
In order to give the simple   basic properties of the 4 dimensional theories we freeze out in turn the degrees of freedom in (\ref{chaKK}). 

Taking a constant dilaton the Gauss-Bonnet term drops out, since in 4 dimensions it is a topological invariant, and we are left with a modified Einstein-Maxwell  theory,
\be
S_{eff}=\int_{\M_4}d^4 x \, \sqrt{-g}  \left\{R-2\Lambda -\quarter F^2 +\frac{3\alpha}{16}  \left[\left(F_{\mu\nu} F^{\mu\nu}  \right)^2-2 F_\mu^\nu F_\nu^\lambda F_\lambda^\kappa F_\kappa^\mu \right] \right\}-S_{int}
\ee
Black hole solutions to this modified Einstein-Maxwell theory and for $\Lambda=0$ have been studied (partially numerically) \cite{chahoyssen2} and they are corrected Reissner-Nordstrom solutions.
For flat spacetime in particular, ie setting $g_{\mu\nu}=\eta_{\mu\nu}$, we get a non-linear version of Maxwell's theory which reads (see the nice paper by Kerner \cite{chakerner} whose notation we follow here),
\be
S_{eff}=\int_{\M_4}d^4 x \, \sqrt{-g} \left\{-\quarter F^2 +\frac{3\alpha}{16}  \left[ \left( F_{\mu\nu} F^{\mu\nu}  \right)^2-2 F_\mu^\nu F_\nu^\lambda F_\lambda^\kappa F_\kappa^\mu \right]\right\} 
\ee
with field equations ($\alpha=\frac{3\gamma}{16 e^2}$),
\ba
\partial_\lambda \left(F^{\lambda\rho} -\frac{3\gamma}{2 e^2} F^2 F^{\lambda \rho}\right)+\frac{3\gamma}{e^2}\partial_\lambda (F_{\mu\nu} F^{\lambda\mu} F^{\rho \nu})&=&0,\nonumber\\
\partial_{[\mu}F_{\lambda \rho]}&=&0
\ea
We see that the usual Maxwell equations are corrected by non-linear terms with coupling $\gamma$. It is interesting to recast everything with respect to the electric and magnetic field, $\vec{E}, \vec{B}$. We then get,
\ba
\label{chamaxwell}
div \vec{E} &=&-\frac{3\gamma}{e^2} \vec{B}\centerdot \vec{grad}(\vec{E}\centerdot \vec{B})\nonumber \\
\vec{rot}(\vec{B}) &=& \frac{\partial \vec{E}}{\partial t}+\frac{3\gamma}{e^2}\left[ \vec{B}\frac{\partial(\vec{E}\centerdot\vec{B})}{\partial t} -\vec{E}\wedge \vec{grad}(\vec{E}\centerdot \vec{B})\right]\nonumber \\
div(\vec{B}) &=& 0 \nonumber \\
\vec{rot}(\vec{E}) &=& -\frac{\partial \vec{B}}{\partial t}
\ea
In this form we can make two obvious remarks. First of all whenever the electric and magnetic field are perpendicular to each other higher order terms drop out and hence usual EM solutions are unchanged. This holds in particular for electromagnetic wave solutions. However note that since we loose linearity one can no longer necessarilly superimpose electromagnetic wave solutions if they are not perpendicular to each other. Furthermore, we can define an induced charge density and current,
\ba
\rho_{ind}=-\frac{3\gamma}{e^2}\vec{B}\centerdot\vec{grad}(\vec{E}\centerdot\vec{B})\nonumber \\
\vec{j}_{ind}=\frac{3\gamma}{e^2}\left[\vec{B} \frac{\partial(\vec{E}\centerdot\vec{B})}{\partial t} - \vec{E}\wedge \vec{grad}(\vec{E}\centerdot \vec{B}) \right] 
\ea
that simulate the higher order terms and verify the continuity equation, $\frac{\partial \rho_{ind}}{\partial t}+div (\vec{j_{ind}})=0.$

Let us now in turn freeze the vector field strength $F_{\mu\nu}=0$. Then we obtain a second order scalar-tensor theory that has been studied quite a lot recently \cite{chadav}. Numerical black hole solutions to such theories were discussed early on in \cite{chawin}. It is important to note that in this case the Gauss-Bonnet 4-dimensional scalar is no longer redundant for the field equations and plays the role of the mediator in-between Einstein gravity and the scalar sector. This is true in whatever frame we choose to go. For example in the Einstein frame, in other words even when the scalar does not couple to linear order with gravity we have that it does so with the Gauss-Bonnet term,
\be
\label{chascaten}
S_{eff}=\int_{\M_4}d^4 x \, \sqrt{-g} \left\{R-(\nabla \phi)^2-2\Lambda e^{\gamma \phi}+\alpha e^{\delta \phi}\hat{G} \right\}
\ee
where $\delta, \gamma$ are some specific couplings depending on the dimensional reduction. Such models are higher order corrected dark energy-quintessence models that are used to evoke late-time acceleration of the universe. The essential point here is that although to leading order they do not affect solar system constaints once one includes the Gauss-Bonnet term this is no longer true and their solar system constraints can be rather stringent \cite{chaamend1}, \cite{chaamend2}. Most stringent constraints arise from light time delay which is calculated from the Cassini spacecraft \cite{chacasini}. For cosmological constraints which are far weaker one can consult \cite{chaamend2}. 
An interesting alternative has been put forward recently \cite{chasolar} in the case of a higher order generalisation of Brans-Dicke theory. It was there shown that the combined effect of the higher order corrections and the scalar sector can reduce or even eliminate the constraint on the Brans-Dicke parameter $\omega$ by imposing particular higher order coupling functions. 

\section{Concluding remarks and open problems}

In these short lectures we saw some of the basic properties of Lovelock theory. Some cosmological applications were given for codimension 1 and 2 braneworlds. There are a large number of open problems in these theories and certain results/issues which we have not treated here. One of the aims of this closing section is to list and comment on some of these open problems. 

Let us start with exact solutions (check out the lecture notes of N Obers for black holes in Einstein gravity). 
For definiteness we studied only the static ones here, a Taub-NUT version of these has been found in \cite{chaman}.
Stationary metrics (akin to the Kerr black hole for 4 dimensional Einstein gravity) have not been found despite efforts. Another important solution which is missing 
is that of the black string (see \cite{chakob} for a perturbative treatment). Although this solution is trivially found in GR in Lovelock 
theory this is not the case. The reason is that Ricci flat solutions in 4 dimensions (as is the 4 dimensional black hole in the 5 dimensional black string) are not vacuum solutions to Lovelock's equations. In fact the bulk Weyl tensor  also contributes to the field equations (unlike in Einstein's equations!) for Lovelock order  $k\geq 2$ \cite{charoy}, \cite{chazam} ie, once we switch on the Gauss-Bonnet invariant. Therefore a 4 dimensional Ricci flat solution which is non-conformally flat will not solve the Lovelock equations (see the nice analysis of \cite{chakastor}). The absence of such simple solutions is maybe an accident but maybe it is also questioning the relevance of the black string type of solutions which are already questionable \cite{chalafla}... At the same time if such an exact solution was found for Lovelock theory it would without doubt be genuinely different from its Einstein version. What could we then say of its stability? Another generic  technical problem is that even in the absence of a bare cosmological constant in the action we always  expect one in the Lovelock solutions due to the presence of multiple  branches which have no flat Einstein theory limit! Thus the candidate metric for resolution must be written in an anzatz suitable for a cosmological constant solution even in the absence of a bare cosmological constant! We know from studies in Einstein theory that beyond a certain symmetry a cosmological constant spoils integrability \cite{charuth}, \cite{chalanglois}. One way around this is to start by looking for solutions at special cases as the Chern-Simons case. Furthermore additional technical problems, due to the absence of Ricci flatness can arise if we simply translate Einstein results. For example,  the diagonal Weyl anzatz in  \cite{chareal}, \cite{charuth} is a priori no longer true for $k>2$ for in order to obtain it we use the fact that the background is Ricci flat{\footnote{I thank Robin Zegers for pointing this out}}! 

Another subject that deserves further attention is that of higher codimension braneworlds. We saw that in the context of Lovelock theory one could in principle define higher codimension braneworlds with Dirac distributions. Only recently \cite{chacod4} did we obtain the first example of a codimension 4 defect having only a removable Dirac singularity. In particular the case of codimension 2 and its cosmology has yet to be elucidated. In this review we explained how one could obtain the exact solutions describing the maximally symmetric branes \cite{chaantonis}. To what extend does Lovelock theory permit us to recover Einstein gravity on the brane \cite{chabos}? Our understanding from \cite{chazeg} is that Lovelock gravity gives us induced Einstein terms on the brane but not Einstein gravity on the brane at all scales. In other words similarily to DGP, only up to some crossover scale, do we expect gravity on a codimension 2 brane to be 4-dimensional. Localised 4 dimensional gravity will occur only when there exists a localised 4-dimensional zero mode graviton in the gravitational spectrum of perturbations. Therefore clearly, what is missing is a clear cut way of developping perturbation theory in Lovelock gravity. We firmly believe that beyond the Gauss-Bonnet term one has to use differential form formalism which we highlighted here. Only then will we know the gravitational spectrum of higher codimension braneworlds and their 4 dimensional phenomenology. We must point out that perturbations of Einstein-Gauss-Bonnet black holes has been carried out in \cite{chaargen} and causality issues have emerged in the context of the adS/CFT correspondance and EGB planar black hole perturbations \cite{chashenker}. 

We hope that with this manuscript we have communicated the fact that Lovelock theory is a technically challenging, interesting, well-motivated and exciting subject of research with numerous open problems that await resolution.

\section*{Acknowledgements}
It is a great pleasure to thank, L Amendola, S C Davis, J-F Dufaux, G Kofinas, A H Padilla, A Papazoglou, R Z Zegers for past and actual collaboration on some of the topics raised
here. I also thank Nemanja Kaloper for many interesting and critical comments concerning numerous issues on Lovelock gravity. I thank Robin Zegers for taking the time and reading through the manuscript and Renaud Parentani for discussions on gravitational instantons. Last but not
least I thank the organisers of the Aegean black hole school for giving me the opportunity to participate and give this set of lectures.

\end{document}